\documentclass[11pt]{article}

\usepackage[a4paper,margin=25mm]{geometry}
\usepackage{graphicx}
\usepackage{multirow}
\usepackage{amsmath,amssymb,amsfonts}
\usepackage{amsthm}
\usepackage{mathrsfs}
\usepackage[title]{appendix}
\usepackage{xcolor}
\usepackage{textcomp}
\usepackage{manyfoot}
\usepackage{booktabs}
\usepackage{algorithm}
\usepackage{algorithmicx}
\usepackage{algpseudocode}
\usepackage{listings}
\usepackage{color,soul}
\usepackage{orcidlink}
\usepackage[numbers,sort&compress]{natbib}
\usepackage{booktabs}
\usepackage{setspace}
\usepackage{doi}
\usepackage{diagbox}
\usepackage{xr-hyper}
\usepackage{hyperref}
\usepackage{rotating}
\usepackage{adjustbox}
\usepackage{caption}
\usepackage[normalem]{ulem}
\usepackage{pdfpages}
\usepackage{enumitem}

\newcommand\rij{\vec{\mathbf r}_{ij}}
\newcommand\rijn{\bigl\lVert \vec{\mathbf r}_{ij}\bigr\rVert}

\theoremstyle{thmstyleone}

\theoremstyle{thmstyletwo}

\theoremstyle{thmstylethree}

\title{Charge density-supervised real-space representations for atomistic energy prediction}

\author{
\begin{tabular}{c}
Heting Liao$^{1,*}$\orcidlink{0000-0002-9939-8367},
Masato Wakeda$^{1,*}$\orcidlink{0000-0002-6377-1318},
Takahito Ohmura$^{1,2}$\orcidlink{0000-0001-7528-566X}
\\[1.0ex]
\parbox{0.9\textwidth}{\centering
\small $^1$Research Center for Structural Materials, National Institute for Materials Science, Tsukuba 305-0047, Japan\\
\small $^2$Department of Materials Science and Engineering, Graduate School of Engineering, Kyushu University, Fukuoka 819-0395, Japan\\[1.0ex]
\small *Corresponding authors: Heting Liao (liao.heting@nims.go.jp), Masato Wakeda (wakeda.masato@nims.go.jp)
}
\end{tabular}
}

\date{}

\begin{document}
\maketitle

\begin{abstract}

Ground-state electron density contains sufficient information to determine ground-state energy, while providing spatial electronic information. This study developed a charge-density-informed graph neural network (hereafter CDGNN) that jointly learns energy and valence electron density from equivariant atomic graphs. This approach incorporates density-supervised real-space probe representations into the global energy regressor, allowing density-related spatial information to contribute to energy prediction. On the QM9 dataset, CDGNN reduced the energy prediction errors relative to an equivariant GNN baseline for interpolation and extrapolation to larger molecules. For small H-containing Fe and Fe--C datasets, CDGNN performed similarly to the baseline, indicating the dataset-dependent benefits of density supervision. CDGNN also accurately predicted the charge density distributions for molecular and Fe-based systems. The attention analysis showed higher weights for probe representations near the H atoms in Fe-based structures. These findings highlight the potential of integrating charge-density information for energy prediction and its task-dependent benefits.

\end{abstract}

\maketitle

\section{Introduction}\label{sec:intro}

Density functional theory (DFT) is one of the most popular methods for material computation and has been widely used to predict the properties of a broad range of materials~\cite{Hafner2006, Hafner2008, Sholl2009}.
However, DFT is limited by its high computational requirements, making its application at large spatial scales difficult.
In recent years, data-driven strategies, such as machine learning-based approaches, have achieved DFT-level accuracy in energy and force prediction at a much lower computational cost~\cite{Schleder2019, Mishin2021, Karniadakis2021, Mortazavi2025}. Therefore, these approaches enable large-scale atomic simulations with near-first-principles accuracy and have attracted considerable attention.
However, because training machine-learning models usually requires high-quality first-principles reference data, which are expensive to generate, they are often trained on relatively limited datasets that may not fully cover the relevant chemical and configurational space~\cite{Butler2018, Mishin2021, Kulichenko2024, Karniadakis2021}. This can lead to unreliable predictions for structures far from the training distribution, making extrapolation a central challenge in machine learning-based atomistic modeling~\cite{Mishin2021}.

The performance of machine learning approaches strongly depends on the training database. A well-trained model can achieve high accuracy for configurations included in or similar to the training set (i.e., interpolation), whereas prediction reliability generally decreases for configurations outside the training set (i.e., extrapolation)~\cite{Omee2024, Shimakawa2024, Xu2021, Yang2023}. Recent studies have suggested that graph neural networks (GNNs) provide better extrapolation than conventional multilayer perceptrons (MLPs) in certain tasks~\cite{Xu2021,Yang2023,Kang2024}. One possible explanation for this is that graph-based architectures and features can encode task-relevant information, facilitating extrapolation~\cite{Xu2021}. In atomistic systems, structures can be naturally represented as graphs, with atoms as nodes and interatomic connections within a cutoff radius as edges~\cite{Scarselli2009,Kearnes2016,Xie2018}. 
Moreover, most atomistic GNNs are based on the message-passing neural network (MPNN) framework~\cite{Gilmer2017}, in which neighboring nodes exchange information and update their representations according to the local atomic environment. Through multiple message-passing layers, information can propagate beyond the immediate local environment, allowing semi-local structural information to be incorporated into learned atomic representations~\cite{Batatia2025}. Message passing has been widely used to predict molecular and material properties~\cite{Gilmer2017,Chen2019,Reiser2022,Xie2018,Schutt2018,Schutt2021,Batzner2022,Banik2023,Louis2020}. In addition, message passing has been suggested to aid in extrapolating GNN interatomic potentials to unseen atomic configurations~\cite{Kang2024}.

In recent years, various GNN architectures have been developed for predicting the physical properties of atomistic systems, ranging from invariant graph models such as CGCNN~\cite{Xie2018}, SchNet~\cite{Schutt2018}, and MEGNet~\cite{Chen2019} to geometry-aware models that explicitly incorporate angular or directional information such as DimeNet~\cite{Gasteiger2020} and ALIGNN~\cite{Choudhary2021}. More recently, equivariant GNNs, including PaiNN~\cite{Schutt2021}, NequIP~\cite{Batzner2022}, and MACE~\cite{Batatia2022}, have emerged as a notable class of models that incorporate orientation-dependent geometric information in a symmetry-consistent manner, thereby enabling a more expressive representation of local atomic environments. Electron density prediction is a representative example because it is a scalar field whose spatial distribution is strongly coupled to the underlying atomic geometry. Successful 3D real-space electron density prediction was demonstrated using DeepDFT~\cite{Jorgensen2022}, which is based on PaiNN~\cite{Schutt2021}, and ChargE3Net~\cite{Koker2024}, which employs an E(3)-equivariant graph neural network.

An important strategy for material-property prediction is the incorporation of physical information into a model~\cite{Butler2018,Mishin2021,Karniadakis2021,Musil2021}. The approach discussed in Ref.~\cite{Mishin2021} combined physics-based functional forms with machine learning. That model predicted the parameters of the physical model that varied with the local atomic environment rather than directly predicting physical quantities from atomic structures. More broadly, physically meaningful information can be incorporated into the learning process as input, supervisory signals, constraints, or intermediate representations, thereby forming a broader class of physically informed machine learning models~\cite{Karniadakis2021,Musil2021}. An example is the introduction of electron density information into energy prediction.
According to the Hohenberg--Kohn theorems~\cite{Hohenberg1964}, the ground-state electron density determines the ground-state energy, making it a key descriptor of the energies, forces, and other derived quantities. Therefore, machine learning models that incorporate electron density information are expected to be more capable of generalization.
For example, Tsubaki and Mizoguchi~\cite{Tsubaki2020} introduced a quantum deep-field model that uses atom-centered Gaussian-type orbital representations to construct electron-density-related intermediate features for molecular energy prediction. The model improved extrapolation from small to large molecules. Del Rio et al.~\cite{delRio2023} developed an ML-DFT emulator that uses predicted electron-density descriptors together with atomic fingerprints to predict energies, forces, and stresses, thereby improving the accuracy and transferability beyond the training domain. Density is represented by learned atom-centered Gaussian basis functions. More recently, Kumar et al.~\cite{Kumar2026} used electron-density fields as physics-aware descriptors to predict material properties. The 3D charge-density grids were compressed using a convolutional autoencoder and used to predict mechanical and thermodynamic properties, demonstrating their potential as transferable electronic structural representations. These studies demonstrate the potential of electron density information as a physically meaningful representation for property prediction.

Building on previous studies, we investigated a different method of incorporating electron density information into atomistic GNNs by jointly learning the electron density and energy from atomic structures. We introduce a charge-density-informed graph neural network (CDGNN) based on an equivariant GNN architecture, in which charge-density-related features are learned together with structural representations and used as additional physical information for energy prediction. Here, the charge density refers to the valence electron density obtained from DFT calculations. The central idea is to use charge density as an additional supervision signal for GNN representation learning, helping the model learn the relationships among atomic geometry, local electronic environments, and energy. The energy and charge density prediction capabilities of CDGNN were evaluated using the QM9 dataset~\cite{Ramakrishnan2014,Ruddigkeit2012} and the H-embedded Fe and Fe--C datasets developed in this study. An attention mechanism~\cite{Velikovi2018, Louis2020, Schmidt2021} was incorporated to identify the spatial charge density information relevant to the energy prediction. Through joint energy--density learning and attention analysis, we examined how charge-density information affects energy prediction and how the model utilizes charge-density information in the prediction process.

\section{Results}\label{sec:results}

\subsection{Charge density-informed graph neural network}
\label{sec:framework}

\begin{figure*}[tp]
\centering
\includegraphics[width=1.0\textwidth]{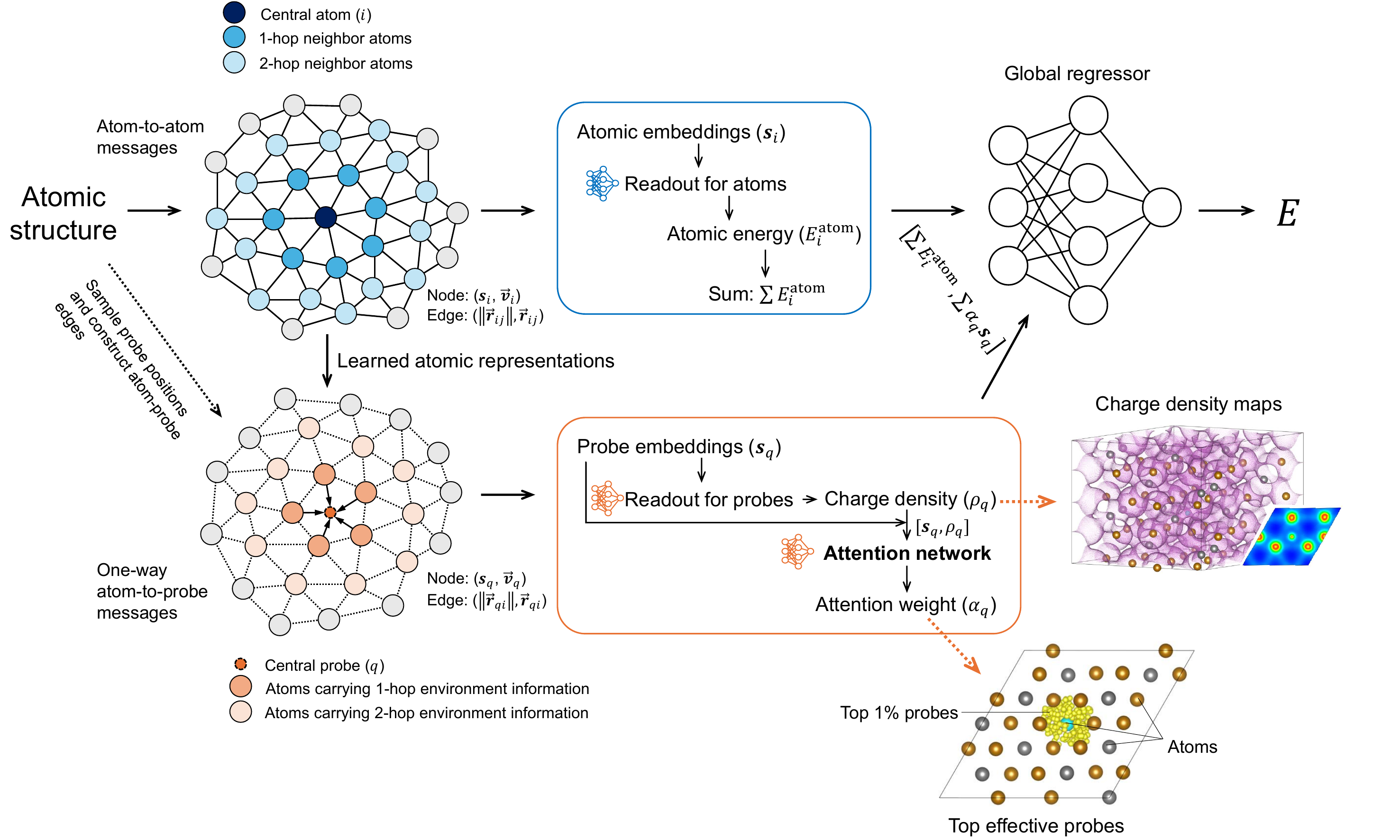}
\caption{Overview of the CDGNN model architecture. The backbone comprises an atom graph and an atom--probe graph. The atom--probe graph is constructed from the atomic structure, and the probe nodes receive atomic representations through one-way atom-to-probe message passing.
The CDGNN model consists of energy branch, charge-density branch, and a global regressor that integrates learned density and energy features to obtain the final total energy prediction. The attention network is used to filter important density probes and suppress noise. The density branch can output charge density maps and model-defined effective probes in 3D real space.}
\label{fig:framework}
\end{figure*}

We developed a CDGNN model based on equivariant message-passing networks~\cite{Gilmer2017, Schutt2021, Jorgensen2022}.
The overall architecture is illustrated in Fig.~\ref{fig:framework}. The CDGNN model contains an atomwise energy branch, a charge-density branch, and a global energy regressor that integrates density features with atomwise energy predictions to produce the final predicted energy. Both branches share a PaiNN-type backbone that encodes local three-dimensional atomic environments using invariant scalar and equivariant vector features. In principle, this framework could be extended to other equivariant network backbones.

The input geometric information for the model included the atomic number, xyz coordinates of each atom and probe, and the unit cell vectors of the crystal structure. A PaiNN-based atomic graph neural network was used to aggregate atomwise information. In the atomic graph, each atom is represented by a vertex, and edges are drawn between vertices if their distance is less than a constant cutoff. These edges were formed under periodic boundary conditions. The atomwise energy predictor maps each final atomic representation to an atomic energy contribution, and the atom-graph energy output is obtained by summing these contributions. 
In parallel, an atom--probe graph following the probe-node strategy of DeepDFT~\cite{Jorgensen2022}, also based on PaiNN, is used to learn charge density information on three-dimensional grid points. A finite number of probe nodes are inserted at the sampled grid positions during training. At these probe positions, the model predicts the charge density from the surrounding atomic structure using DFT-calculated valence electron densities as reference values. In addition, to avoid artificial probe nodes affecting the atomic representations, the probe nodes receive only unidirectional messages from the atom nodes within the cutoff sphere. The hidden probe states are initialized to zero and updated using the aggregated incoming atom-to-probe messages. 

In the charge density branch, an attention network~\cite{Velikovi2018, Louis2020, Schmidt2021} is jointly trained with the other modules. For each probe, the attention network predicts the attention score from the probe representation and the predicted charge density. These scores are normalized across all probes with the same configuration to obtain the attention weights. The
weights were then used to pool the probe representations into a configuration-level density-informed representation, which was then passed to the subsequent global energy regressor.
The attention network assigns larger weights to probes that are more relevant to energy prediction, while suppressing less informative or noisy contributions. Further details regarding the attention network are provided in Section~\ref{sec:Attent}. Finally, the atomwise energy predictions and pooled probe representations were combined using a global regressor for the final energy prediction.

Notably, CDGNN jointly learns the mappings from the atomic structure to the charge density and energy. Charge-density supervision guides the model in learning local electronic features, which are also used in energy prediction. Thus, the two mappings are coupled through shared learned representations. After training, these probe representations were inferred from their atomic structures. No reference charge density was required for energy prediction.

\subsection{Datasets}

\begin{figure*}[bp]
\centering
\includegraphics[width=1.0\textwidth]{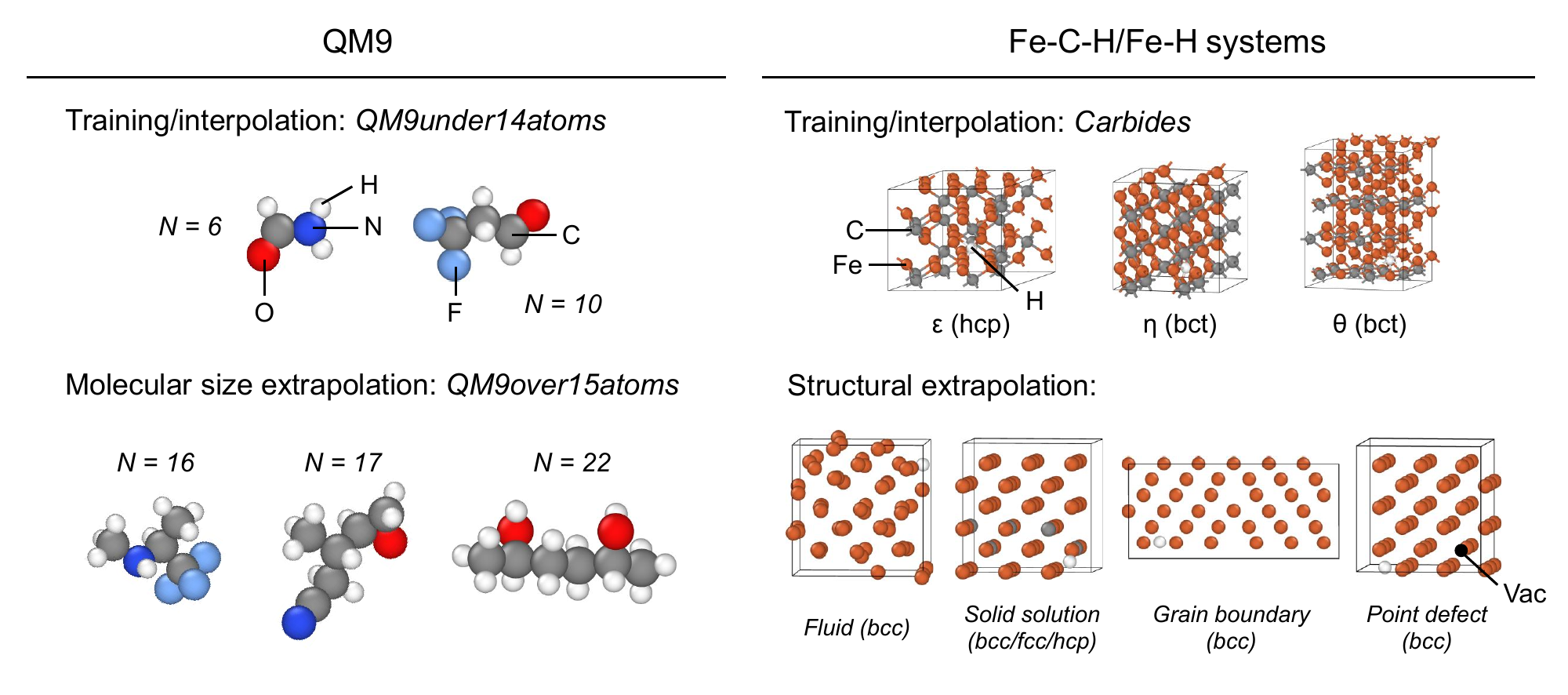}
\caption{Representative structures for interpolation and extrapolation in the QM9 and Fe-based datasets.}
\label{fig:dataset}
\end{figure*}

\begin{figure*}[tbp]
\centering
\includegraphics[width=1.0\textwidth]{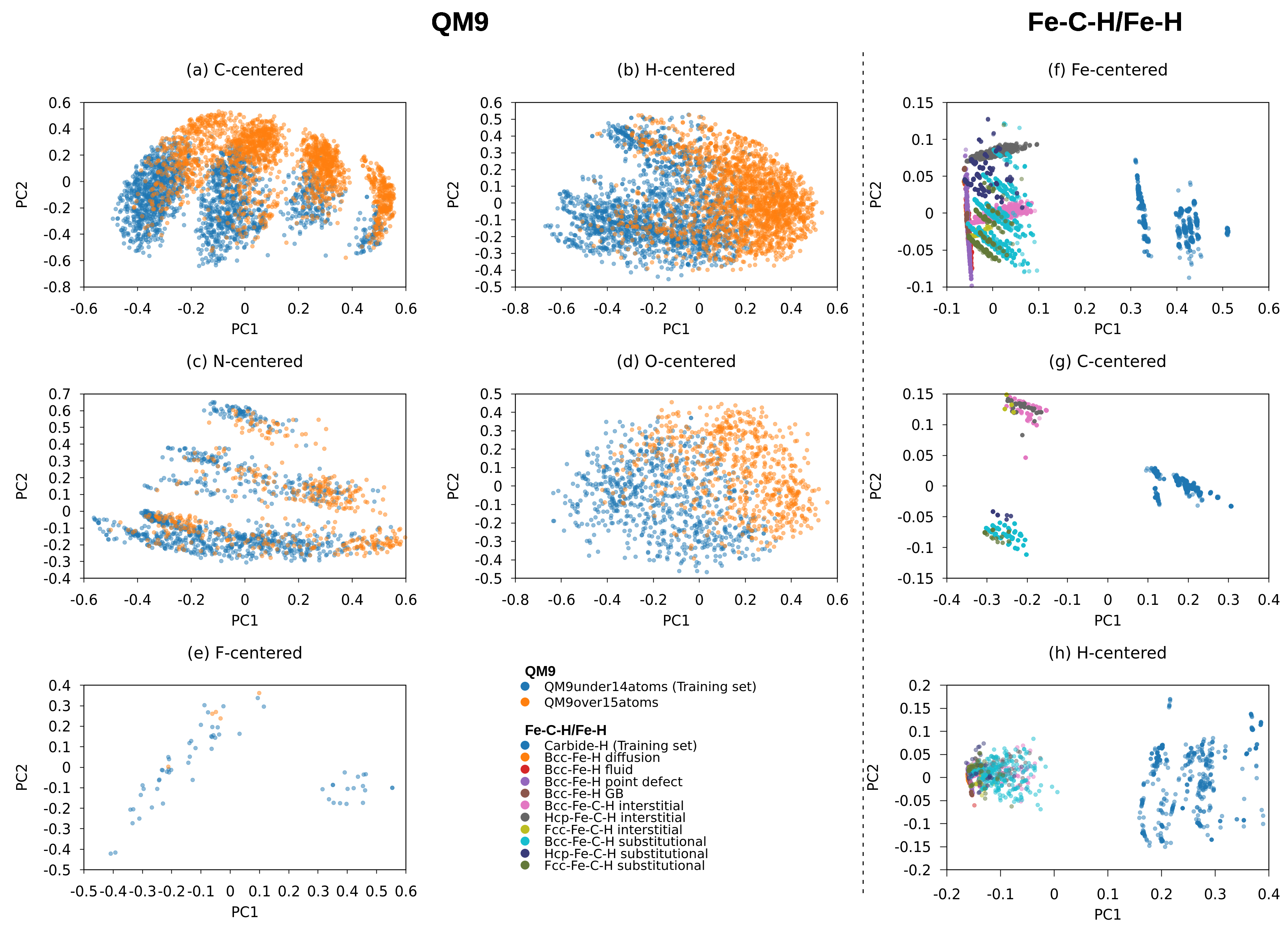}
\caption{PCA visualization of local atomic environments for the QM9 and Fe-based datasets.}
\label{fig:pca}
\end{figure*}
 
Molecular systems from the QM9 dataset and H-embedded Fe-based systems were used to evaluate the CDGNN framework.
The QM9 dataset contains 134k stable small organic molecules comprising C, H, O, N, and F, and their atomic structures were obtained from Ref.~\cite{Ramakrishnan2014}. We used these to compile a charge-density database for training the CDGNN model. The construction of the QM9 charge-density dataset is detailed in Section~\ref{sec:dataset}. By extrapolating the energy predictions from Ref.~\cite{Tsubaki2020}, we separated the samples into the \textit{QM9under14atoms} and \textit{QM9over15atoms} datasets. The \textit{QM9under14atoms} dataset contained 15862 molecules with no more than 14 atoms and was used for the training and interpolation tests. The \textit{QM9over15atoms} dataset contained the remaining 118203 molecules for the extrapolation test. Examples of the structures included in the training and extrapolation sets are shown in Fig.~\ref{fig:dataset}. The prediction target was chosen as the total energy of each molecule. Figure~\ref{fig:pca} compares the local atomic environments across datasets using principal component analysis (PCA) of smooth overlap of atomic positions (SOAP) descriptors. The two QM9 subsets largely overlap in the PCA plots, with slight differences in their distributions.

Several small datasets of H-embedded Fe-based systems with distinct structures, containing a total of 1620 samples, were built. The datasets included carbides with and without strains, strained bcc-Fe, fluid bcc-Fe configurations, bcc-Fe with point defects, $\Sigma3\langle 110 \rangle$ grain boundary (GB) structures in bcc-Fe, and solid solutions in body-centered cubic (bcc), face-centered cubic (fcc), and hexagonal close-packed (hcp) lattices with substitutional and interstitial solutes. Each configuration contains one H atom. The training dataset constructed from strained and unstrained carbides (hereafter referred to as \textit{carbide--H}) contained 657 atomic configurations of hcp-$\varepsilon$, bct-$\eta$, and bct-$\theta$ carbides. Here, ``bct'' denotes body-centered tetragonal. The carbide--H dataset was used for training because it includes all chemical species considered and spans diverse structural environments, providing a broad training domain for transferability tests.
The remaining datasets were used for the extrapolation testing (Fig.~\ref{fig:dataset}).
The details of the DFT calculations and atomic configuration selection are provided in Section~\ref{sec:dataset} and Supplementary Information~\ref{sec:dataset-Fe-extra}.
The hydrogen segregation energy defined in Eq.~\ref{eq:e_seg} was used as the energy prediction target for Fe-based systems. A previous DFT study~\cite{Liao2026} showed that the hydrogen segregation energies in Fe carbides are associated with the local charge environment around the segregation sites. This provides a suitable test of whether the charge density information incorporated into CDGNN can contribute to segregation energy prediction. See Fig.~\ref{fig:pca} for PCA visualization for Fe-based datasets. Unlike those in the QM9 dataset, the Fe–C–H/Fe–H training and test sets show no overlap, suggesting greater differences in their local environments. These differences may make generalization more difficult for the Fe-based datasets.


\subsection{Model configurations and training}

The CDGNN models were trained on the \textit{QM9under14atoms} and \textit{carbide--H} datasets using a feature size of $F$ = 128, cutoff radius of $r_{c}$ = 6.0~\AA, and $T$ = 3 interaction steps. Additional training details, including the loss function and optimization settings, are provided in Section~\ref{sec:training_settings}.
An energy-only PaiNN-type equivariant GNN shared the same PaiNN atomic encoder and atomwise energy-prediction branch as CDGNN. 
Here, we used a PaiNN-type GNN that uses only geometric information as a baseline model (i.e., GNN baseline).
The training parameters of the baseline models matched those used for the CDGNN. The baseline GNN has fewer trainable parameters when the same feature dimensions are used. To acquire a more comparable parameter count, we constructed a larger baseline model with $F$ = 206, for which the total number of trainable parameters approximately matched that of CDGNN (i.e., GNN baseline (increased $F$)). The baseline models were trained using only the energy loss term in Eq.~(\ref{eq:loss}).
A CDGNN variant without an attention network was also considered (i.e., CDGNN (w/o AttentNet)). In this variant, the attention network was replaced by average pooling over all probe representations to obtain a single-density representation. The model parameters are presented in Table~\ref{tab:para}.

All models were trained for $10^7$ and $10^6$ steps on the \textit{QM9under14atoms} and \textit{carbide--H} datasets, respectively, using three independent $8:1:1$ splits for training, validation, and interpolation testing. Unless otherwise stated, the prediction errors shown below were averaged over the three models trained using independent data splits. For each model, the checkpoint with the lowest validation error was selected for evaluation. The corresponding training curves are provided in the Supplementary Information \ref{sec:training_curves}. No significant overfitting was observed during training.

\begin{table*}[t]
\caption{Number of parameters of various GNN models.}\label{tab:para}%
\begin{tabular*}{\textwidth}{@{\extracolsep\fill}lccc}
\toprule
Model & Feature size & Interaction step &  Number of parameters \\
\midrule
GNN baseline					&	128	&	3	&	611329\\
GNN baseline	 (increased $F$)	&	206	&	3	&	1530169\\
CDGNN 						&	128	&	3	&	1537412\\
CDGNN (w/o AttentNet)		&	128	&	3	&	1520643\\
\bottomrule
\end{tabular*}
\end{table*}


\subsection{Energy predictions}
\label{sec:energy_prediction}

The energy prediction error was evaluated after normalization by the number of atoms in each structure. The per-atom energy mean absolute error (MAE) and root mean square error (RMSE) are defined as:
\begin{equation}
	\text{MAE}_\text{energy} 
	=
	\frac{1}{M}\sum_{m=1}^{M}
	\left|
	\frac{\hat{E}_m - E_m}{N_m}
	\right|,
\end{equation}
\begin{equation}
	\text{RMSE}_\text{energy}
	=
	\sqrt{
	\frac{1}{M}\sum_{m=1}^{M}
	\left(
	\frac{\hat{E}_m - E_m}{N_m}
	\right)^2
	}.
\end{equation}
Here, $\hat{E}_m$ and $E_m$ are the predicted and reference DFT-calculated energies of structure $m$, respectively; $N_m$ is the number of atoms in the structure, and $M$ is the number of structures in the dataset. As described in Section~\ref{sec:framework}, CDGNN predicts the energy using atomwise structural features and probe-based electronic features. During prediction, randomly distributed virtual probes were introduced around the atomic structure to obtain the probe-based electronic representation. Because the probe positions were sampled randomly, an insufficient number of probes may have introduced variations in the predicted energy. Therefore, 2000 probes were used for each prediction as a compromise between sampling stability and computational cost. To quantify the variation arising from random probe sampling, we performed 1000 repeated predictions for a 129-atom Fe--H structure, with the probe positions independently resampled for each prediction. The resulting energies showed a mean absolute deviation of 0.04~meV/atom and a standard deviation of 0.06~meV/atom. To further suppress this sampling-induced variation, every CDGNN energy value reported in this study was obtained by averaging 10 independent predictions, each using 2000 newly sampled probe positions.

\begin{figure*}[tb]
\centering
\includegraphics[width=\textwidth]{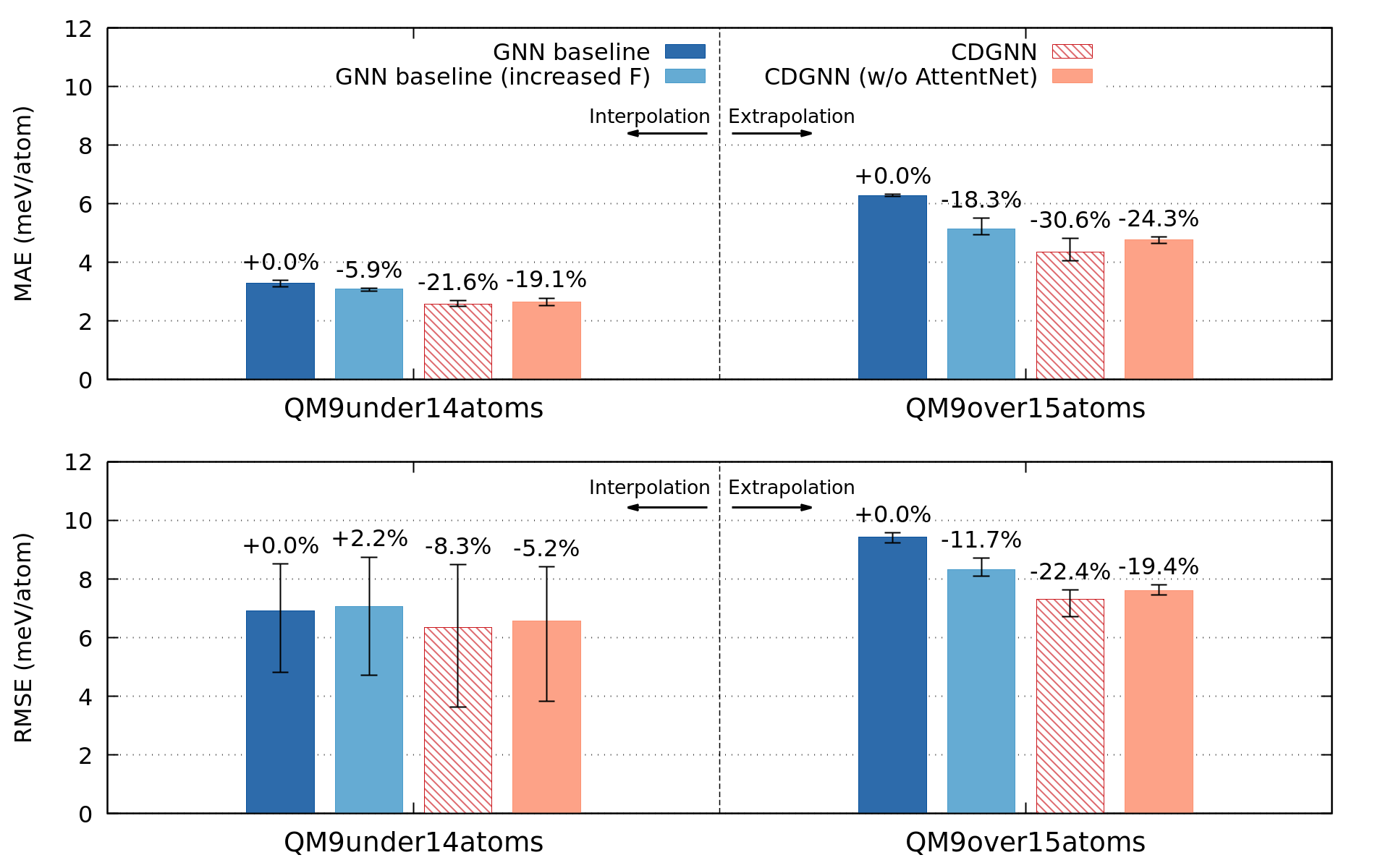}
\caption{Energy MAE and RMSE of various models trained on the QM9 dataset for predicting total energy.
The pattern-filled blocks represent the proposed CDGNN model. The \textit{QM9under14atoms} dataset was used for training, whereas the \textit{QM9over15atoms} dataset was used for extrapolation testing. The \textit{QM9under14atoms} results shown here correspond to the interpolation test split, which was not included in the training process. Percentages indicate error relative to the GNN baseline.
}
\label{fig:error-qm9}
\end{figure*}

The calculated energy errors of the models trained on the \textit{QM9under14atoms} dataset are presented in Fig.~\ref{fig:error-qm9}. All equivariant GNN-based models achieved high accuracy for total-energy prediction (Fig.~\ref{fig:error-qm9}). In the size extrapolation test from smaller to larger molecules, CDGNN yielded the lowest errors among the tested models (Fig.~\ref{fig:error-qm9}). Compared with the GNN baseline, CDGNN reduced both the MAE and RMSE, indicating better size-extrapolation performance for larger molecules. 
As shown in Fig.~\ref{fig:error-qm9}, the geometry-only baseline with an increased feature size also improves the extrapolation performance, indicating that increasing the model capacity can be beneficial for chemically diverse datasets such as QM9. Even in comparison with this parameter-matched baseline with a similar number of trainable parameters, CDGNN still provided the lowest MAE and RMSE. 
The CDGNN w/o AttenNet model performs better than the GNN baseline but shows a slightly higher error rate than the CDGNN. This suggests that the improvement may come from both the probe-based density representations and the attention mechanism. The attention network also provides an additional means to examine the spatial distribution of the probe contributions to the energy readout, as discussed in Section~\ref{sec:important_e_info}.


\begin{figure*}[tb]
\centering
\includegraphics[width=\textwidth]{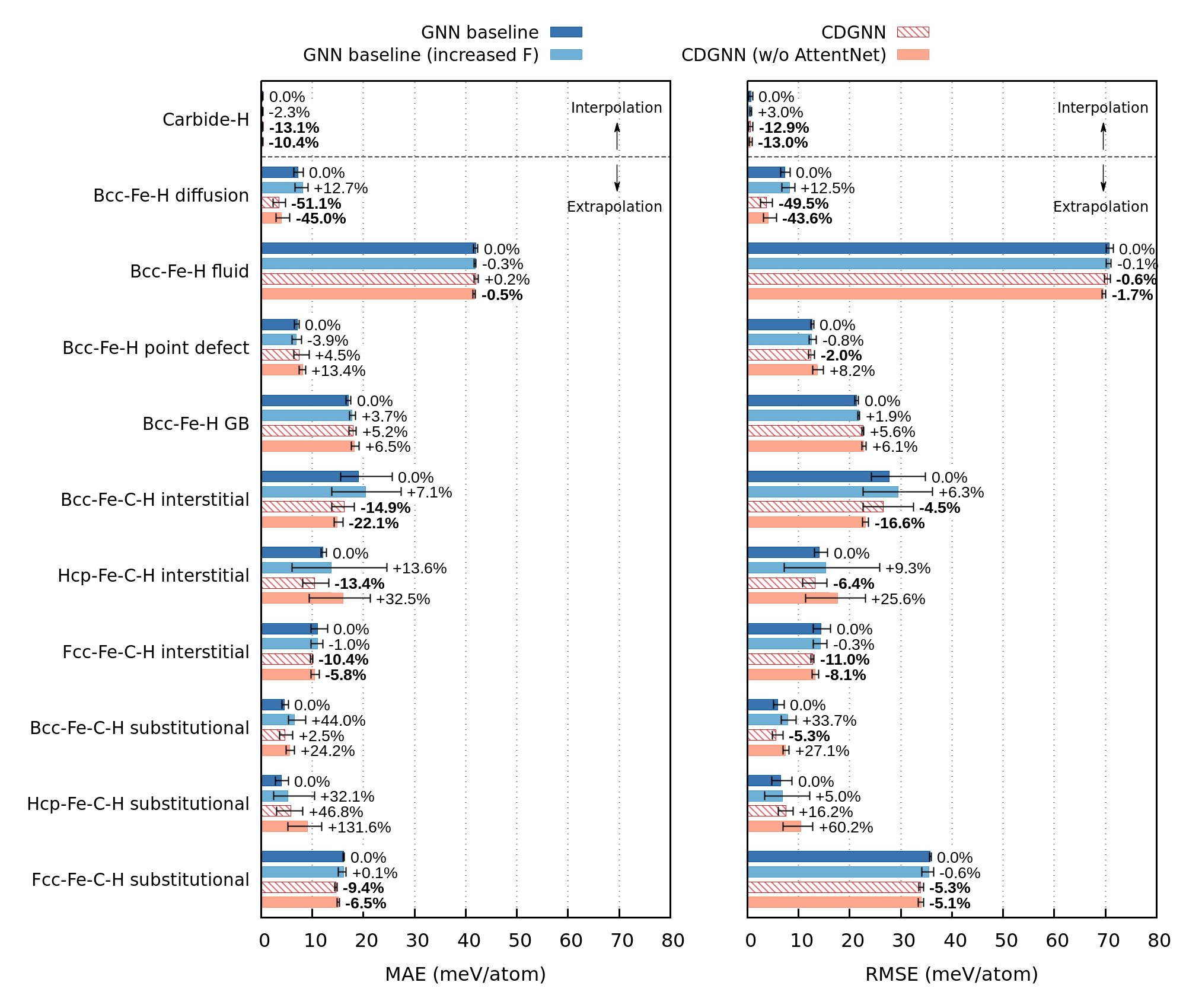}
\caption{Energy MAE and RMSE of various models trained on the \textit{carbide--H} dataset for predicting H segregation energy. The pattern-filled blocks represent the proposed CDGNN model. The \textit{carbide--H} results shown here correspond to the interpolation test split, which was not included in the training process. Bold percentages indicate error reductions relative to the GNN baseline for CDGNN and CDGNN w/o AttentNet.}
\label{fig:rmse-Fe-H}
\end{figure*}

The energy errors of the models trained and tested on the Fe--C--H/Fe--H dataset were also evaluated. In contrast to the results obtained for QM9, both the baseline and CDGNN models, including their variants, generally performed poorly in the extrapolation tests. Increasing the feature size $F$ of the baseline model can either improve or degrade the extrapolation performance, depending on the dataset. The corresponding results are shown in Fig.~\ref{fig:rmse-Fe-H} and Supplementary Information~\ref{sec:energy_prediction}. In the interpolation test on the \textit{carbide--H} dataset, all models showed high accuracy (Fig.~\ref{fig:rmse-Fe-H}). For most of the extrapolation datasets containing structurally distinct configurations, none of the models reproduced either the energy values or their trends (Figs.~\ref{fig:e_extra_1} and \ref{fig:e_extra_2}).
More reasonable predictions were obtained for other datasets such as \textit{bcc-Fe--H diffusion} and \textit{bcc/hcp-Fe--C--H substitutional}, possibly because these configurations were more similar to the bct and hcp carbide structures included in the training set. The corresponding errors are presented in Figs.~\ref{fig:rmse-Fe-H}. For the datasets showing relatively good extrapolation performance, as indicated by RMSE $<$ 20~meV/atom in Fig.~\ref{fig:rmse-Fe-H}, removing the attention network from CDGNN reduces prediction accuracy.

Because the prediction target for Fe-based systems is the hydrogen segregation energy, which does not scale with the system size, the baseline model was modified by introducing a global regression module similar to that used in CDGNN. The module comprises a small MLP that takes the sum of the atomwise energy contributions and mean-pooled atomic representation as input and produces the final structure-level energy prediction. This design was motivated by previous work~\cite{Schweidtmann2023}, which showed that the choice of the pooling function should reflect the size dependence of the target property. As shown in Supplementary Information~\ref{sec:energy_prediction}, introducing the global regressor can either improve or degrade prediction accuracy depending on the specific extrapolation task, without showing an overall advantage over CDGNN. These results suggest that modifying the energy readout architecture alone does not consistently improve extrapolation performance.

To examine the joint learning mechanism of CDGNN, we initialized both the atomic graph backbone and the atomwise energy head with parameters from a a pretrained energy branch of CDGNN, including the atomic graph and atomwise energy head, and kept them frozen while training the probe branch, attention network, and global energy regressor. As shown in Supplementary Information~\ref{sec:freeze}, CDGNN with the frozen energy branch achieved prediction accuracy comparable to that of the GNN baseline but substantially lower than that of the end-to-end trained CDGNN for an extrapolation sample from the \textit{bcc-Fe--H diffusion} dataset. These results suggest that training the additional density-related components alone is insufficient to reproduce the improved extrapolation performance of CDGNN, highlighting the importance of end-to-end joint optimization.

Consequently, the results suggest that CDGNN can outperform the baseline for molecular size extrapolation in QM9, whereas the improved performance for structural extrapolation in Fe--C--H/Fe--H systems cannot be guaranteed and strongly depends on the dataset and specific extrapolation task. In both the QM9 and Fe-based datasets, CDGNN still demonstrated better performance than the baseline model with a similar number of parameters. For a relatively successful extrapolation, the attention network may also contribute to improving the energy prediction accuracy.


\subsection{Charge density predictions}
\label{sec:charge_prediction}

The CDGNN model can also predict charge density distributions in three-dimensional real space. To evaluate the charge density prediction accuracy, the relative density error for each structure is defined as~\cite{Grisafi2018, Fabrizio2019, Jorgensen2022}
\begin{equation}
\epsilon_\text{density} = \frac{\int \left| \hat{\rho}(\vec{\mathbf r}) - \rho(\vec{\mathbf r}) \right| dV}{\int \left| \rho(\vec{\mathbf r}) \right| dV} \times 100\%,
\end{equation}
where $\vec{\mathbf r}$ denotes the spatial position and $\hat{\rho}$ and $\rho$ are the predicted and reference charge density values, respectively. In practice, the integrals are evaluated on a charge-density grid. The dataset-level error was obtained by averaging the relative errors over all structures, $\epsilon_\text{density}^\text{avg} = \frac{1}{M} \sum^{M}_{m=1} \epsilon_\text{density}^m$, such that each structure contributed equally to the reported value. The results of the QM9 and Fe--C--H/Fe--H datasets are presented in Tables~\ref{tab:abs_error_qm9} and \ref{tab:abs_error}, respectively. To reduce the computational cost, a charge-density evaluation of the \textit{QM9over15atoms} extrapolation dataset was performed using 1000 randomly selected configurations.

\begin{table*}[tb]
\caption{Average relative errors (\%) in charge-density predictions of various models for the QM9 dataset relative to the DFT results. Uncertainties are given as standard deviations.}\label{tab:abs_error_qm9}%
\begin{tabular*}{\textwidth}{@{\extracolsep\fill}lcc}
\toprule
Dataset   & CDGNN & CDGNN  \\
& &   (w/o AttentNet) \\
\midrule
\textit{QM9under14atoms}					& 	0.39 $\pm$ 0.01 & 0.43 $\pm$ 0.01\\
1000 configurations in \textit{QM9over15atoms}	&	0.55 $\pm$ 0.00 & 0.62 $\pm$ 0.02\\
\bottomrule
\end{tabular*}
\end{table*}

\begin{table*}[tb]
\caption{Average relative errors (\%) in charge-density predictions of various models for the Fe--C--H/Fe--H datasets relative to the DFT results. Uncertainties are given as standard deviations.}\label{tab:abs_error}%
\begin{tabular*}{\textwidth}{@{\extracolsep\fill}lccc}
\toprule
Dataset & DeepDFT  & CDGNN & CDGNN  \\
& &  & (w/o AttentNet) \\
\midrule
\textit{Carbide--H}				&	0.05 & 	$0.07 \pm 0.03$ &$0.09 \pm 0.02$\\
\midrule
\textit{Bcc-Fe--H diffusion}			&	0.37 &	$0.69 \pm 0.16$& $0.86 \pm 0.24$\\
\textit{Bcc-Fe--H fluid}			&	0.83 & 	$1.30 \pm 0.22$ &$1.20 \pm 0.11$\\
\textit{Bcc-Fe--H point defect}		&	0.70 & 	$1.28 \pm 0.19$ &$1.16 \pm 0.11$\\
\textit{Bcc-Fe--H GB}				&	0.65 & 	$0.90 \pm 0.14$ &$0.93 \pm 0.15$\\
\textit{Bcc-Fe--C--H interstitial}		&	1.24 & 	$1.23 \pm 0.29$ &$1.11 \pm 0.10$\\
\textit{Hcp-Fe--C--H interstitial}		&	1.89 & 	$1.35 \pm 0.22$ &$1.59 \pm 0.24$\\
\textit{Fcc-Fe--C--H interstitial}		&	0.78 & 	$1.37 \pm 0.20$ &$1.20 \pm 0.17$\\
\textit{Bcc-Fe--C--H substitutional}	&	0.64 & 	$1.17 \pm 0.22$ &$1.06 \pm 0.13$\\
\textit{Hcp-Fe--C--H substitutional}	&	0.83 & 	$0.96 \pm 0.15$ &$1.31 \pm 0.41$\\
\textit{Fcc-Fe--C--H substitutional}	&	0.87 & 	$1.52 \pm 0.19$ &$1.33 \pm 0.16$\\
\bottomrule
\end{tabular*}
\end{table*}

As shown in Table~\ref{tab:abs_error_qm9}, the charge density prediction errors of CDGNN models are comparable to those reported for equivariant DeepDFT (0.27\%)~\cite{Jorgensen2022} and ChargE3Net (0.20\%)~\cite{Koker2024}. These studies also used structures derived from QM9, although the vacuum spacing around the molecules and data split differed from those used in this study. Although CDGNN was trained only on smaller molecules, the charge-density errors for larger molecules remained within a similar range.
We also compared the charge density predictions of CDGNN with those of an equivariant DeepDFT model trained using the same \textit{carbide--H} dataset for Fe-based systems. The training settings of the DeepDFT model are provided in the Supplementary Information~\ref{sec:deepdft}. As shown in Table~\ref{tab:abs_error}, both CDGNN and DeepDFT retained reasonable charge density prediction accuracy in the interpolation and extrapolation tests. CDGNN usually produces slightly larger errors than DeepDFT, possibly reflecting the additional constraint introduced by jointly optimizing the energy and charge density in the multitask framework.
The extrapolation errors for Fe-based systems were larger than the interpolation errors. This differs from the behavior observed for QM9 and may result from larger differences in the local chemical environments between the training and extrapolation datasets. A similar contrast between the QM9 and Fe-based datasets was observed for energy prediction.


\subsection{Spatial distribution of model attention in energy prediction}
\label{sec:important_e_info}

\begin{figure*}[tbp]
\centering
\includegraphics[width=\textwidth]{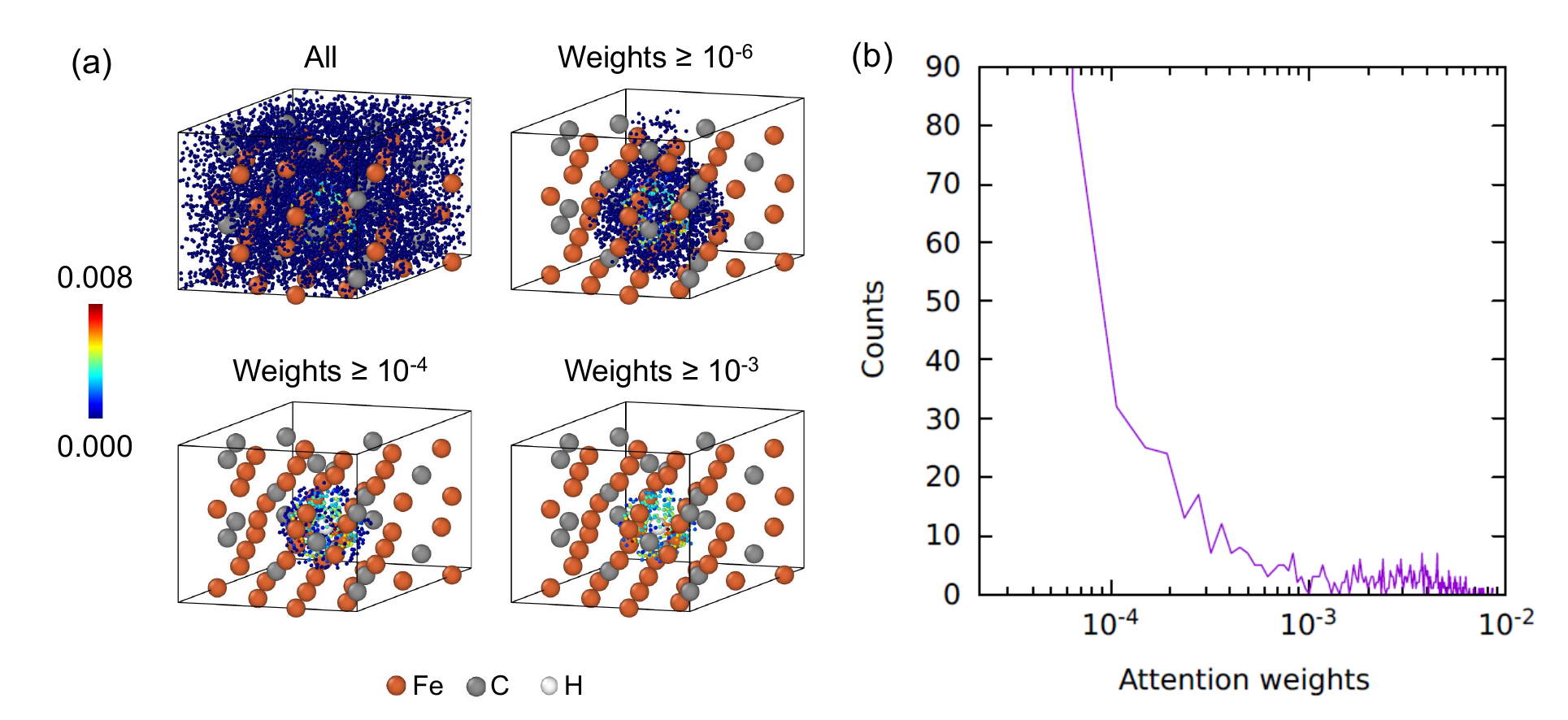}
\caption{Attention weights predicted by the attention network in the CDGNN model for a H-embedded $\varepsilon$-carbide configuration in the interpolation set. (a) Atomic configurations with charge density probes are colored according to their attention weights. The large and small spheres are atoms and probes, respectively. (b) Probe counts shown as a function of attention weight. The 8000 random probes in each configuration were collected from a single run.}
\label{fig:atten_sample}
\end{figure*}

\begin{figure*}[tp]
\centering
\includegraphics[width=\textwidth]{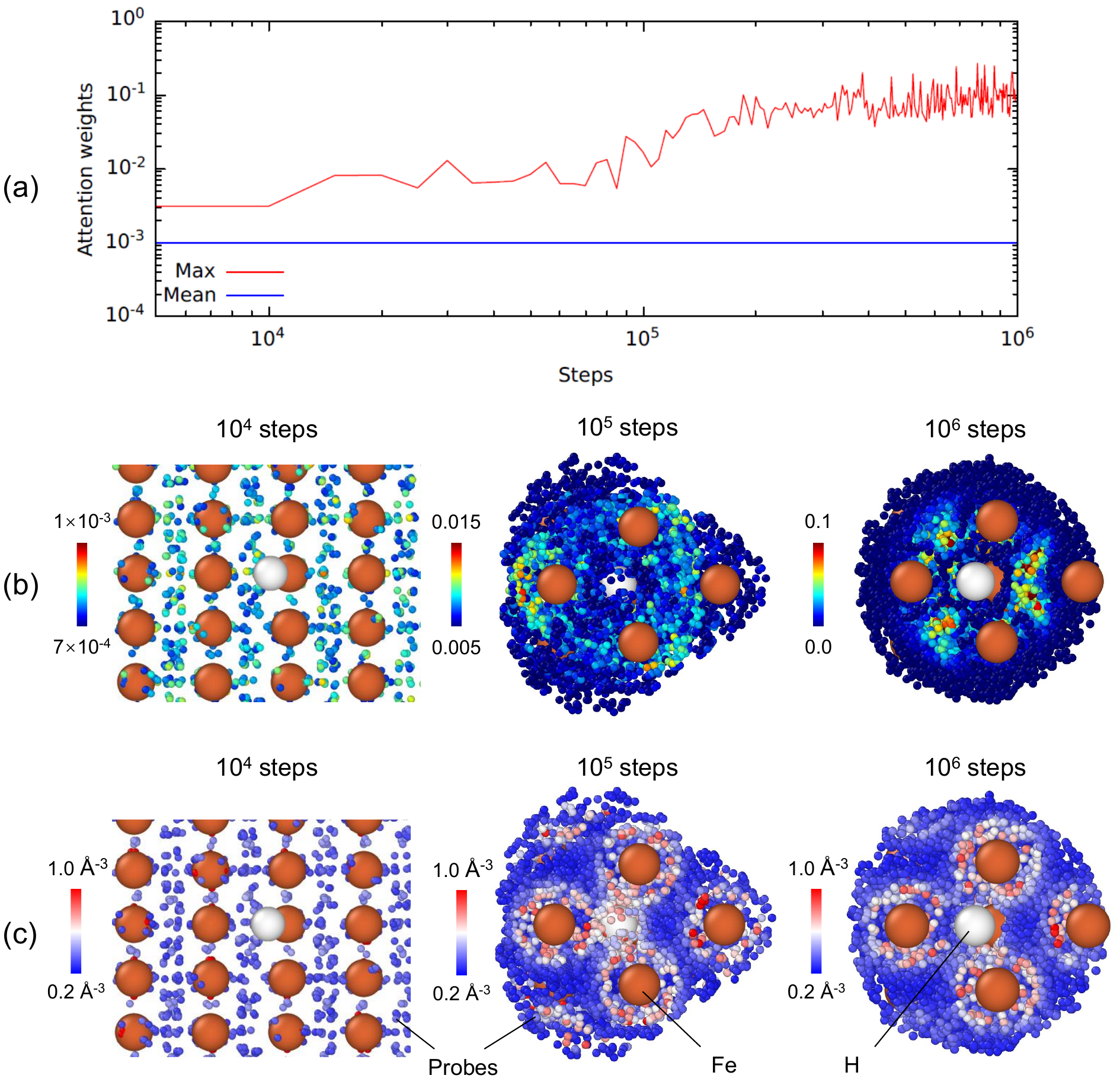}
\caption{Prediction of the attention network in the CDGNN model: (a) Attention weights during training. (b) Top-ranked probes based on their attention weights and (c) model predicted charge density on probes in the bcc-Fe structure with a H atom embedded at the tetrahedral site. The data in (b, c) were obtained using three CDGNN models trained with $10^4$, $10^5$, and $10^6$ steps, and the slices on the (110) plane were shown. Only the top 4\% probes among $2\times10^5$ sample probes are shown. The large orange and white spheres represent Fe and H atoms, respectively. The small spheres are charge density probes that are colored according to their attention weights.}
\label{fig:atten_short}
\end{figure*}

The attention network is a learnable module in the CDGNN framework that assigns an attention weight to each probe during energy prediction. These weights indicate the relative importance assigned by the model to different probe-based charge-density representations. For each probe, the attention network first output an unnormalized score based on the probe representation and predicted charge density. The scores were then normalized using a softmax function over all probes in the same structure to obtain the attention weights. Further details on the attention network are provided in Section~\ref{sec:Attent}. To visualize the spatial distribution of the probes emphasized by CDGNN, the probes sampled during energy prediction were colored according to their attention weights, as shown in Fig.~\ref{fig:atten_sample}. This analysis was performed using the Fe--C--H/Fe--H datasets.
As shown in Fig.~\ref{fig:atten_sample}, the probes were randomly distributed throughout the supercell, whereas probes with higher attention weights were preferentially located near the H atom. This pattern was observed in both the interpolation and extrapolation datasets. Previous DFT studies~\cite{Liao2026, Zhang2021, Liu2025} have shown that hydrogen segregation energies in Fe carbide and bcc-Fe systems are associated with the local electronic environment around the segregation sites. The spatial distribution of the learned attention weights is qualitatively consistent with the physical picture.

Figure~\ref{fig:atten_short} shows the evolution of the attention weights and the predicted charge density during CDGNN training. As the number of training steps increases, the maximum attention weight first increases and then gradually plateaus (Fig.~\ref{fig:atten_short}(a)). To examine the spatial evolution of the learned attention, we plotted the highest-weighted probes obtained from multiple random probe samplings using CDGNN models saved at different training steps.
As shown in Fig.~\ref{fig:atten_short}(b), the highly weighted probes gradually became concentrated around the H atom during training. At $10^5$ training steps, the top-ranked probes were already concentrated around H, but their spatial distribution was less clearly defined. After $10^6$ steps, they became more localized in several anisotropic regions near the H atom, mainly in the interstitial space between the H and neighboring Fe atoms. This evolution suggests that the attention weights are not simply determined by the distance from the H atom. As shown in Fig.~\ref{fig:atten_short}(c), the predicted charge density values changed only slightly between $10^5$ and $10^6$ training steps, whereas the locations of the top-ranked probes changed substantially. This further suggests that the attention weights are not solely determined by the magnitude of the predicted charge density. Overall, CDGNN assigned higher attention weights to the selected probe representations near the H atom when constructing the density-informed representation used for energy prediction. This behavior is qualitatively consistent with the relevance of the local H--metal electronic environment to the hydrogen segregation energetics.


\subsection{Runtime and scalability}

\begin{figure}[t] 
\centering
\includegraphics[width=.7\textwidth]{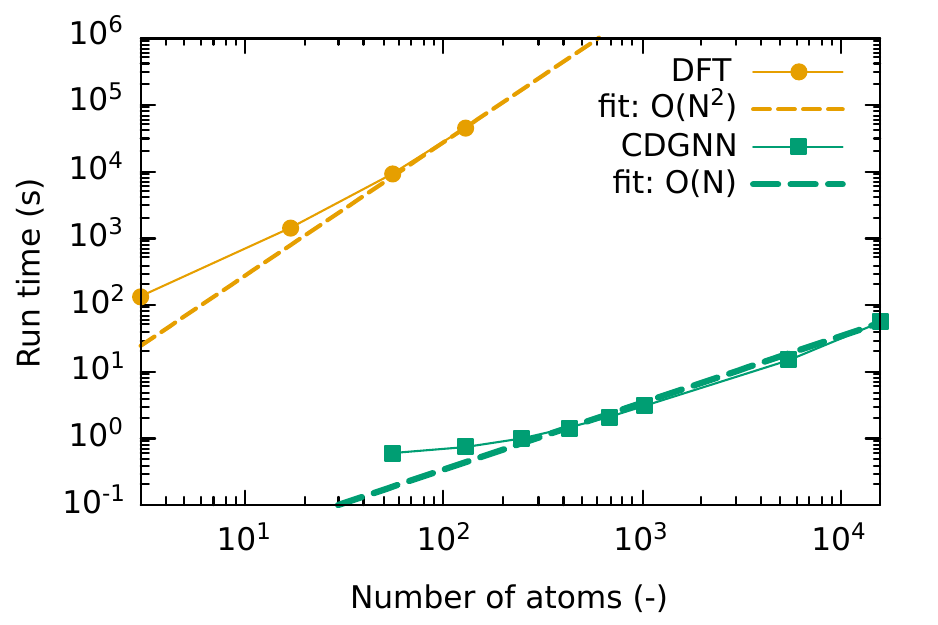}
\caption{Runtime of DFT calculations and CDGNN energy predictions. The DFT calculations and CDGNN predictions were performed on an Intel Xeon Silver 4510 CPU and an NVIDIA GeForce RTX 4090 GPU, respectively. The reported CDGNN runtime corresponds to a single energy prediction with charge-density output disabled. A fixed number of 2000 probe samples was used for each CDGNN prediction. Dashed lines indicate reference scaling trends.}
\label{fig:runtime}
\end{figure}

The CDGNN runtime was evaluated for systems with an increasing number of atoms and compared to that of DFT calculations. Bcc-Fe supercells of up to $30\times30\times30$, each containing one H atom, were used for the energy calculations. CDGNN predictions were performed using the PyTorch-based implementation developed in this study. For each energy prediction, 2000 probe points were randomly sampled. For the energy-prediction runtime evaluation, the charge-density output was disabled, whereas the probe representations required for the energy readout were retained.
DFT calculations were performed using the Vienna Ab initio Simulation Package (VASP)~\cite{Kresse1996}. The DFT runtime was defined as the time required for the static calculation of a fixed atomic configuration to achieve electronic self-consistency within a single ionic step. The electronic convergence criterion was set to $10^{-5}$~eV. All other settings were identical to those used in constructing the Fe--C--H/Fe--H datasets (Section~\ref{sec:dataset}).
Figure~\ref{fig:runtime} shows the runtime scaling of CDGNN and DFT calculations with increasing system size. The CDGNN runtime increased approximately linearly for large systems and remained substantially lower than the DFT runtime under the computational settings used here.
For charge-density prediction, the CDGNN runtime is comparable to that of DeepDFT~\cite{Jorgensen2022}.


\section{Discussion}\label{sec:discussion}

This paper presents an equivariant neural network model, CDGNN, for predicting charge-density-informed energy. During training, the model jointly learns energies and charge densities from atomic structures, allowing the incorporation of electronic information into the learned representations. During prediction, only the atomic structure is required as the input, from which both the energy and three-dimensional charge density distribution can be predicted.
The final energy prediction combines information from the atomic structure and charge-density branch. The probe representations are constructed from the atomic representations through one-way atom-to-probe message passing, and are incorporated into the energy prediction at the final readout. An attention network was designed to suppress excessive noise in the charge density information and identify probe representations that are more relevant to energy prediction. The resulting attention weights provide an interpretable indication of the charge density spatial regions that contribute more strongly to the model output. Jointly learning the atomic structure, electron density, and energy, therefore, provides spatially resolved electronic information for interpreting energy predictions and may also improve model generalization.

These results show that on QM9, the CDGNN outperformed the geometry-only equivariant GNN when predicting the energies of larger molecules using a model trained only on smaller molecules. However, the same advantage was not observed for the Fe--C--H/Fe--H dataset.
This difference may arise from the characteristics of the datasets and extrapolation tasks. In QM9, the extrapolation from smaller to larger molecules mainly tests the size transfer within a relatively continuous chemical space. Although the molecular size increases, the test structures contain elements and local bond environments. In contrast, the Fe-based tests involve a much larger distribution shift. The model was trained on carbide--H structures, whereas the extrapolation datasets included strained and fluid-like bcc-Fe and bcc-Fe with point defects, $\Sigma3\langle110\rangle$ grain boundaries, and substitutional or interstitial solid solutions in bcc, fcc, and hcp lattices. These structures differ in their host lattices, defect types, local coordination, and chemical environments. The lack of a consistent advantage of CDGNN in these tests therefore suggests that the benefit of the density branch depends strongly on both the dataset and the type of extrapolation.

One possibility is that the benefit of charge-density supervision becomes more evident when the training data cover a broader and more continuous range of structural and chemical environments. Under these conditions, density supervision may help the model learn representations that are useful across different configurations. However, this requires further validation using larger and more diverse datasets.
Furthermore, a comparison between CDGNN models with and without the attention network suggests that attention may improve energy prediction accuracy in some cases, although its effect depends on the dataset. The attention network also provides a means to examine the spatial distribution of attention weights over probe representations, which may help relate the learned representations to the local electronic environment.

In the present implementation, CDGNN learns the energy and charge density jointly from the beginning of the training. This allows probe representations to develop under charge-density supervision, while the model is optimized for energy prediction. However, the two targets have different forms: the charge density is a spatially resolved field defined at many probe points, whereas the energy is a single scalar for the entire structure. This difference leaves room to further optimize the training strategy, including the training schedule and relative weighting of the two loss terms. For example, pretraining the density branch may allow probe representations to learn the local electronic environment before being coupled to energy prediction. Gradually adjusting the weights of loss during training may also help balance density and energy learning. These strategies may improve the incorporation of density-related probe representations into final energy predictions.
In the current architecture, probe representations are constructed from learned atomic representations, providing a one-way coupling from the atomic graph to the probe graph. Future work may explore stronger coupling by feeding the density-related probe representations back to update the atomic representations.

The proposed CDGNN model has several limitations. First, when the goal is only energy prediction, CDGNN has a higher computational cost than the geometry-only GNN. The training data required additional charge-density calculations, and the atom--probe branch increased both the training time and memory usage. Therefore, the additional cost may not always be justified for interpolation tasks in which a simpler energy-only model performs well.
Second, energy prediction for large systems may become sensitive to the number of sampled probes. With a fixed number of probes, the sampling density per atom decreased as the system size increased, which may have reduced the resolution of the probe-based representation. Increasing the number of probes can improve the sampling resolution, but also increases the memory usage and computational cost.
Third, the current weighted loss jointly supervises energy and charge density but does not directly establish whether density-related representations contain physically meaningful information relevant to energy prediction. Further evaluations and ablation tests are required to examine this relationship.

The advantages of CDGNN are its flexible framework and the spatial information incorporated into the energy prediction. From a framework perspective, CDGNN does not rely on manually designed descriptors to encode the atomic structure and electron density information separately. Instead, atomwise and probe-based representations are learned within a common equivariant graph neural network framework. Although additional training data and validation are required for each new class of material, this design makes the model readily adaptable to systems with different elements and chemical environments.
A distinct capability of CDGNN is that the spatially resolved probe representations learned under electron density supervision are incorporated into the energy readout. Therefore, energy prediction uses real-space probe information in addition to atomwise representation. Although these probe features and their attention weights do not directly represent the physical energy contributions, they provide a spatial indication of the regions emphasized by the model during energy prediction. Further validation of the physically defined electronic quantities and material properties may provide useful spatial information for analyzing charge redistribution and identifying candidate regions associated with energetic changes. 
Beyond energy prediction, CDGNN could be extended to predict other properties using electron-density supervision, such as molecular dipole moments. With additional training targets and suitable extensions, the framework could also be explored for response properties such as polarizability. Functional materials, including dielectric and ferroelectric materials, are promising systems to investigate because charge redistribution and polarization play important roles in their properties. In these systems, electron-density supervision may help the model learn spatial electronic features relevant to the target properties. Exploring these applications could help clarify when this supervision offers the greatest benefit.

\section{Methods}\label{sec:methods}

\subsection{Framework of charge density-informed graph neural network}\label{sec:GNN}

\subsubsection{Graph construction}

The GNN backbone follows the neural message-passing framework of Gilmer et al.~\cite{Gilmer2017}. In this framework, each atom is represented as a node with a hidden state. The node representations are iteratively updated by exchanging messages with neighboring atoms. Following the PaiNN~\cite{Schutt2021} architecture, rotation equivariance is achieved by augmenting the hidden state of each node with a vector component. Throughout this paper, we use bold symbols to represent quantities with multiple components, and arrows to indicate vectors in real space. Given the atomic coordinates $\{\vec{\mathbf r}_i\}$, we define two rotation-equivariant graphs in 3D space as follows:


\bigskip

\noindent \textbf{Atom graph.}
For each atom node $i$ in the atom graph, the model assigns a scalar--vector state $(\mathbf{s}_i,\vec{\mathbf{v}}_i)$. The scalar channel is initialized by mapping the atomic number $Z_i$ to a learnable $H$-dimensional feature vector $(\mathbf{s}_i\in\mathbb{R}^{H})$, and is subsequently updated by message passing to encode the local atomic environment. The vector channel is initialized as zero and evolves into $H$ three-dimensional vector features $(\vec{\mathbf{v}}_i\in\mathbb{R}^{3\times H})$ that carry directional geometric information and transform equivariantly under spatial rotations.
The atom graph is constructed with vertices corresponding to atoms, and the edges between atoms $i$ and $j$ are created if the interatomic distance $ \bigl\lVert \vec{\mathbf r}_{ij} \bigr\rVert < r_\text{cut}$, where $r_\text{cut}$ is the cutoff distance.

The residual structure of message passing and update is used.
The scalar state is updated by a sum over messages from neighboring atoms~\cite{Schutt2018, Schutt2021, Jorgensen2022}
\begin{equation}
	\Delta \mathbf s^m_i = \sum_{j} \mathbf \Phi_s (\mathbf s_j) \circ \mathbf W_s(\rijn) f_\text{cut}(\rijn),
	\label{eq:residual1}
\end{equation}
where $\mathbf \Phi_s$ is a two-layer neural network that transforms the neighboring scalar state $\mathbf s_j$, and $\circ$ is element-wise multiplication. The filter $\mathbf W_s$ is rotationally invariant and is constructed from a sinc-like radial basis expansion of the distance~\cite{Gasteiger2020}, $\sin(\frac{n\pi}{r_\text{cut}}\rijn)/\rijn$, with $1\leq n \leq 30$. A cosine cutoff function~\cite{Behler2007}, $f_\text{cut}(\rijn) = \frac{1}{2}[\cos(\pi \rijn / r_\text{cut})+1]$ for $\rijn < r_\text{cut}$ and 0 otherwise, is applied to smoothly damp the messages near the cutoff.

Similarly, the vector state is updated by~\cite{Schutt2021, Jorgensen2022}:
\begin{equation}
	\begin{aligned}
		\Delta \vec{\mathbf v}_i^m = & \sum_{j} \vec{\mathbf v}_j \circ \mathbf \Phi_{vv}(\mathbf s_j) \circ \mathbf W_{vv}(\rijn) f_\text{cut}(\rijn) \\
		&+  \sum_{j} \mathbf \Phi_{sv}(\mathbf s_j) \circ \mathbf W_{sv}(\rijn) f_\text{cut}(\rijn) \frac{\rij}{\rijn},
		\label{eq:residual2}
	\end{aligned}
\end{equation}
where $\mathbf \Phi_{vv}$ and $\mathbf \Phi_{sv}$ denote learnable transformations of the neighboring scalar state $\mathbf s_j$, and $\mathbf W_{vv}$ and $\mathbf W_{sv}$ are rotationally invariant radial filters. $vv$ denotes the vector-to-vector contribution that propagates existing vector features, while $sv$ denotes the scalar-to-vector contribution that generates new vector messages along the edge direction. In Eq.~(\ref{eq:residual2}), the first term propagates the existing equivariant vector features $\vec{\mathbf v}_j$ from neighboring atoms $j$ to atom $i$. The second term introduces new directional information through the unit displacement vector $\rij/\rijn$ along the edge direction. Consequently, these two contributions update the vector state while preserving rotation equivariance.

Furthermore, two additional atomwise update equations are applied to improve expressiveness.
For scalar features, the following equation is used~\cite{Schutt2021}:
\begin{equation}
	\Delta \mathbf s^u_i =
	\mathbf a_{ss} (\mathbf s_i, \bigl\lVert \mathbf V \vec{\mathbf v}_{i}\bigr\rVert)
	+ \mathbf a_{sv} (\mathbf s_i, \bigl\lVert \mathbf V \vec{\mathbf v}_{i}\bigr\rVert)
	\circ
	\langle \mathbf U \vec{\mathbf v}_{i}, \mathbf V \vec{\mathbf v}_{i}\rangle,
	\label{eq:residual3}
\end{equation}
where $(\cdot,\cdot)$ represents the concatenation of features, $\langle \cdot,\cdot\rangle$ denotes the inner product over spatial vector components, and $\mathbf a_{ss}$ and $\mathbf a_{sv}$ are learnable transformations obtained from the scalar state and the norm of transformed vector features. The vector features use the following equation~\cite{Schutt2021}:
\begin{equation}
	\Delta \vec{\mathbf v}^u_i =
	\mathbf a_{vv} (\mathbf s_i, \bigl\lVert \mathbf V \vec{\mathbf v}_{i}\bigr\rVert)
	\circ
	\mathbf U \vec{\mathbf v}_{i},
	\label{eq:residual4}
\end{equation}
which scales a linear combination of equivariant vector features. These updates couple the scalar and vector channels while preserving the invariant/equivariant structure of the representation.

After the last atom-graph interaction step, the final scalar state of each atom is mapped to a scalar atomwise energy contribution $E_i^{\mathrm{atom}}$. The atom-graph energy output is then obtained by summing these atomwise contributions, i.e., $E_{\mathrm{atom}} = \sum_{i=1}^{N} E_i^{\mathrm{atom}}(\mathbf s_i)$.

\bigskip

\noindent \textbf{Atom--probe graph.}
The atom--probe graph uses the same message-passing form as the atom graph in Eqs.~(\ref{eq:residual1}) and (\ref{eq:residual2}), but with one-way messages from atom nodes to probe nodes. In line with previous studies~\cite{Jorgensen2022, Koker2024}, probe nodes are introduced for charge density prediction and only receive
messages from neighboring atomic nodes. The hidden states of the probe vertices are initialized to zero.
For the atom--probe interaction, a gating network~\cite{Jorgensen2022} is used to dynamically combine the previous probe state with the incoming atom-to-probe message:
\begin{equation}
	\mathbf s_q^{\text{new}} =
	\mathbf G_s (\Delta \mathbf s_q^m) \circ \mathbf s_q
	+ (\mathbf1 - \mathbf G_s (\Delta \mathbf s_q^m)) \circ  \Delta \mathbf s_q^m,
\end{equation}
\begin{equation}
	\vec{\mathbf v}_{q}^{\text{new}} =
	\mathbf G_v (\Delta \mathbf s_q^m) \circ \vec{\mathbf v}_{q}
	+ (\mathbf1 - \mathbf G_v (\Delta \mathbf s_q^m)) \circ  \Delta \vec{\mathbf v}_{q}^m.
\end{equation}
Here, $q$ denotes a probe node, $\mathbf s_q$ and $\vec{\mathbf v}_q$ are the current scalar and vector states of the probe node, and $\Delta \mathbf s_q^m$ and $\Delta \vec{\mathbf v}_q^m$ are the scalar and vector messages aggregated from neighboring atom nodes. The functions $\mathbf G_s$ and $\mathbf G_v$ are learnable gates with sigmoid
outputs, with their values between 0 and 1. These gates determine how much of the previous probe state is retained and how much of the incoming atom-to-probe message is incorporated. After this gated atom-to-probe message aggregation, the scalar--vector update in Eqs.~(\ref{eq:residual3}) and (\ref{eq:residual4}) is applied to the probe states. After the final interaction step, the final scalar state of each probe vertex is mapped to a scalar charge-density value $\rho_q$.


\subsubsection{Attention network}
\label{sec:Attent}

To implement attention-based pooling for charge density information in the CDGNN model, we first concatenate each probe state $\mathbf s_q$ output by the atom--probe graph with its corresponding density value $\rho_q$ to form a probe density feature $\mathbf u_q = \bigl(\mathbf s_q, \rho_q\bigr)$ for a single probe. Including $\rho_q$ enables the attention score to depend explicitly on the predicted charge density at the probe position. This probe density feature is then passed through a two-layer attention neural network~\cite{Velikovi2018, Louis2020}
\begin{equation}
	\text{AttentionNet}(\mathbf u_q)=a_q=\mathbf w_2^\top\bigl(\tanh(\mathbf W_1\,\mathbf u_q + \mathbf b_1)\bigr)+b_2,
\end{equation}
where $\mathbf W_1 \in \mathbb{R}^{H\times(H+1)}$, $\mathbf b_1 \in \mathbb{R}^H$, $\mathbf w_2 \in \mathbb{R}^H$, and $b_2 \in \mathbb{R}$ are trainable weights and biases. The attention network uses a tanh activation function for the hidden units and outputs an attention score $a_q$ for each probe.
After scoring all the probes, the resulting scores were normalized to probes of the same configuration using a softmax function to produce the attention weight:
\begin{equation}
\alpha_q = \frac{\exp(a_q)}{\sum_{p=1}^{P}\exp(a_{p})}.
\end{equation}
The probe states are then aggregated into a configuration-level vector using an attention-weighted summation $\mathbf z = \sum_{q=1}^{P} \alpha_q \mathbf s_q$. The resulting representation is invariant to the ordering of the probes within the configuration. This attention mechanism allows the model to focus on probes that carry more relevant density information and potentially suppress less-informative contributions.


\subsubsection{Global energy regressor}

To achieve geometry-to-density/energy mapping and enable CDGNN energy prediction, a two-layer MLP with a SiLU activation function in the hidden layer was adopted at the end of the framework. The atomwise energy sum $\sum_i E_i^\text{atom}$ predicted from the atom graph is concatenated with the attention-pooled probe representation $\mathbf z$ obtained from the atom--probe graph, where the attention weights are computed using both the probe state $\mathbf s_q$ and the predicted density $\rho_q$. This concatenated vector is then passed to the MLP to produce the final CDGNN energy prediction $E_{\text{pred}} = \mathrm{MLP}\left(\sum_i E_i^\text{atom}, \mathbf z\right)$.
The present implementation does not feed density-related features back into the atomic representations. Such strong coupling may be explored in future work.


\subsection{Dataset construction and DFT calculations}\label{sec:dataset}

We compiled a QM9 CHGCAR database using first-principles calculations via VASP~\cite{Kresse1996}. The atomic structures contain 134k stable small organic molecules made up of CHONF obtained from Refs.~\cite{Ramakrishnan2014,Ruddigkeit2012}. The VASP setting is given in Refs.~\cite{Jorgensen2022, jorgensen2022qm9}. Compared to the dataset built in Refs.~\cite{Jorgensen2022, jorgensen2022qm9}, we increased the vacuum layer to at least 8~\AA \ in the supercell around the molecules to decrease the interaction between adjacent molecules. The calculations employed the generalized gradient approximation (GGA with the PBE function)~\cite{Perdew1996}. The interactions between electrons and nuclei were described using the PAW method~\cite{Blochl1994, Kresse1999}. The Brillouin zone was sampled using a gamma-point-only {\it k}-point mesh. A plane-wave cutoff energy of 400~eV was applied. The convergence cutoff for the electronic self-consistency loop was set to $10^{-5}$~eV. We performed only static calculations to obtain the energy and charge densities. 

We also prepared hydrogen‐embedding datasets for the Fe--C--H and Fe--H systems for training, validation, and testing using DFT calculations. The datasets included carbides with and without strain, strained bcc-Fe, fluid bcc-Fe configurations, bcc-Fe with point defects, $\Sigma3$$\langle 110 \rangle$ GB structures in bcc-Fe, and solid solutions in the bcc/fcc/hcp lattices (substitutional and interstitial). The selection details of atomic configurations are provided in Supplementary Information~\ref{sec:dataset-Fe-extra}. A charge density file (CHGCAR) and corresponding energy file were prepared for each structure. The charge density and corresponding energy values were obtained from the same structure. We adopted the hydrogen segregation energy as the predicted target so that the model could directly learn and predict the segregation energy. Each configuration contained a single H atom, and the segregation energy was calculated as follows: 
\begin{equation}
	E_\text{seg} = E(\text{Fe}_x\text{C}_y\text{H}) - E(\text{Fe}_x\text{C}_y) - [ E(\text{Fe}_x\text{H}) -  E(\text{Fe}_x) ],
	\label{eq:e_seg}
\end{equation}
where $E(\text{Fe}_x\text{C}_y\text{H})$ and $E(\text{Fe}_x\text{C}_y)$ are the total energies of the carbide systems with and without a H atom embedded, respectively, while $E(\text{Fe}_x\text{H})$ and $E(\text{Fe}_x)$ are the total energies of the bcc-Fe matrix with and without a H atom embedded, respectively. $E(\text{Fe}_x\text{H})$ is calculated from the structure of the H atom embedded at the tetrahedral site.
Note that $E(\text{Fe}_x\text{C}_y\text{H})$ is obtained using different relaxation strategies depending on the specific case, but $E(\text{Fe}_x\text{C}_y)$, $E(\text{Fe}_x\text{H})$, and $E(\text{Fe}_x)$ are always obtained under 
The DFT calculations were performed using VASP~\cite{Kresse1996}. The calculations employed the framework of the generalized gradient approximation (GGA) in the Perdew--Burke--Ernzerhof (PBE) form~\cite{Perdew1996}. The interactions between electrons and nuclei have been described using the projector-augmented wave (PAW) method~\cite{Blochl1994, Kresse1999}. {\it k}-space integrations were performed using the Monkhorst--Pack~\cite{Monkhorst1976} sampling scheme. The spacing between the {\it k}points in reciprocal space was 0.19~\AA$^{-1}$. A plane-wave cutoff energy of 500~eV was applied, and the geometries were relaxed until the residual forces decreased below 1~meV/\AA. The convergence cutoff for the electronic self-consistency loop was set to $10^{-8}$~eV.
The charge densities were output on a grid with a spacing of approximately 0.07~\AA. Each configuration comprised $10^6$–$10^7$ grid points.

\subsection{Training settings}
\label{sec:training_settings}

The CDGNN model learns joint mappings among the geometry, charge density, and energy. The trainable parameters of the different network components are jointly updated during backpropagation using the following training loss:
\begin{equation}
\begin{split}
    \mathcal{L}
    &= w_\text{density} \cdot \mathcal{L}_\text{density}
    + w_\text{energy} \cdot \mathcal{L}_\text{energy} \\
    &= w_\text{density} \cdot
    \frac{1}{\sum_{b=1}^{B}P_b}
    \sum_{b=1}^{B}\sum_{q=1}^{P_b}
    (\hat{\rho}_{b,q} - \rho_{b,q})^2
    + w_\text{energy} \cdot
    \frac{1}{B}\sum_{b=1}^{B}
    \left(\frac{\hat{E}_b - E_b}{N_b}\right)^2 .
    \label{eq:loss}
\end{split}
\end{equation}
where $\hat{\rho}_{b,q}$ and $\rho_{b,q}$ are the predicted and reference charge density values at probe $q$ of structure $b$, respectively; $\hat{E}_b$ and $E_b$ are the predicted and reference total energies, respectively; $N_b$ and $P_b$ are the numbers of atoms and probes in structure $b$; $B$ is the batch size; $w_\text{density}$ and $w_\text{energy}$ control the relative weights of the two loss terms. Through this joint training, the model learns how the atomic configurations determine the local charge density field and how density-related features contribute to the final energy prediction. Therefore, the charge density target serves as an additional supervision signal. 

For CDGNN models, we used a feature size $F = 128$, cutoff radius $r_{c} = 6.0$~\AA, and interaction steps $T = 3$ during training using the \textit{QM9under14atoms} and \textit{carbide--H} datasets. The CDGNN model and its variants were trained with loss weights of $w_\text{density}$ = 1 and $w_\text{energy}$ = 10. Weight sensitivity is discussed in Supplementary Information~\ref{sec:sensitivity_to_weights}. Owing to the large memory requirements of the charge density files, we employed a rotating pool of 20 atomic configurations and a batch size of two atomic configurations when training the CDGNN model. From each training step, we uniformly sampled 1000 probes from a predefined charge-density grid for each configuration. Numerous training steps were required to capture the entire electronic structure of the training set. The Adam optimizer was used with an initial learning rate $10^{-4}$, which decayed exponentially with $\eta_g = 10^{-4} \times 0.96^{g/10^{5}}$, where $g$ is the gradient step index. For early stopping, the validation set was evaluated every 5000 steps, and the checkpoint with the lowest error was retained. During validation of the CDGNN model, probe sampling drew 5000 probes per configuration.


\section*{Code availability}
The underlying code for this study will be made publicly available upon acceptance of the manuscript.

\section*{Data availability} 
The datasets generated and analyzed during the current study are not publicly available due to institutional data management policies but are available from the corresponding author on reasonable request.

\section*{Author contributions}
H.L. contributed to the conceptualization, methodology, software, data analysis, database creation, and writing of the manuscript.  M.W. contributed to the conceptualization, provided computational resources, and edited the manuscript.
T.O. reviewed the manuscript and acquired funding. All the authors have read and approved the final version of the manuscript. 

\section*{Acknowledgements}
This study was supported by the MEXT Program: Data Creation and Utilization Type Material Research and Development Project, Grant Number JPMXP1122684766, and the Japan Society for the Promotion of Science (JSPS) KAKENHI Grant Number 24H00389. Computations were conducted using a numerical material simulator at the National Institute for Materials (NIMS). The authors thank Dr. Masahiko Demura for insightful discussions and helpful comments.

\section*{Competing interests}
All authors declare no financial or non-financial competing interests.

\bibliographystyle{unsrtnat}
\bibliography{bibliography}

\clearpage
\begin{center}
    \huge\bfseries Supplementary information
\end{center}

\renewcommand\thesection{S\arabic{section}}
\setcounter{section}{0}

\counterwithin{figure}{section}
\counterwithin{table}{section}
\counterwithin{equation}{section}

\renewcommand\thefigure{S\arabic{section}.\arabic{figure}}
\setcounter{figure}{0}
\renewcommand\thetable{S\arabic{section}.\arabic{table}}
\setcounter{table}{0}
\renewcommand\theequation{S\arabic{section}.\arabic{equation}}
\setcounter{equation}{0}


\section{Datasets in Fe--C--H/Fe--H systems}
\label{sec:dataset-Fe-extra}

The following datasets in the Fe--C--H/Fe--H systems were built. To reduce the computational cost, only static calculation were performed for building this dataset. In this study, the \textit{carbide--H} dataset was used for training and interpolation testing, while the remains were used for the extrapolation test.

\begin{itemize}[label=--]
\item \textit{carbide--H} dataset contains 657 atomic configurations of multiple $\varepsilon$-, $\eta$-, and $\theta$-carbides in bcc-Fe with a hydrogen atom embedded at the Voronoi vertices. Each configuration has 68--128 atoms with one hydrogen atom. To minimize the interactions between the periodic images of the hydrogen atom in the DFT calculations, supercells with cell edges larger than 9~\AA\ were used. 
For more details on the carbide structures, refer to Ref.~\cite{Liao2026}. The dataset included configurations with and without geometry relaxation. Additionally, the strained carbide systems and hydrogen migration structures between segregation sites in $\varepsilon$-$\text{Fe}_{2.4}\text{C}$, $\varepsilon$-$\text{Fe}_{3}\text{C}$, and $\theta$-$\text{Fe}_{3}\text{C}$ with and without strain, as considered in Ref.~\cite{Liao2026}, were also included in the dataset. The applied elastic strain was up to 20 \%. The dataset includes hydrogen migration structures in some carbides with and without strain, obtained from climbing image-nudged elastic band (CI-NEB) calculations~\cite{Henkelman2000Improved, Henkelman2000A}. For more details on the carbide dataset, see Ref.~\cite{Liao2026}.

\item \textit{bcc-Fe--H diffusion} dataset contains 194 atomic configurations, including 128 Fe atoms and one H atom. 
A $4\times4\times4$ supercell was used for the DFT calculation. For each migration route, the initial and final positions of the hydrogen atom are the tetrahedral and octahedral sites of the bcc-Fe matrix, and 12 intermediate images are created by interpolating at equal intervals along the path. These structures were considered with less than 1\% elastic strain in the x, y, and z-directions. Note that the initial bcc-Fe matrix was under fully structural relaxation and the obtained lattice constant was 2.82~\AA. The intermediate images were then created using the relaxed Fe matrix, and further calculations were performed with the geometry fixed.

\item \textit{Fe--H fluids} dataset was made by performing molecular dynamics calculation using a $3a_\text{bcc}\times3a_\text{bcc}\times3a_\text{bcc}$ ($a_\text{bcc}=2.86$ \AA) Fe supercell via a BPNN potential for the Fe--C system developed by Meng et al.~\cite{Meng2024} and Large-scale Atomic/Molecular Massively Parallel Simulator (LAMMPS) code~\cite{Thompson2022}. The molecular dynamics calculation was performed for $2\times10^5$ steps at 1800 K, close to the melting point of steel. After that, 33 structures were randomly selected from the snapshots with the introduction of interstitial hydrogen atoms as the initial structures for static DFT energy and charge density calculations.

\item \textit{bcc-Fe--H point defect} dataset was create using a $3a_\text{bcc}\times3a_\text{bcc}\times3a_\text{bcc}$ Fe supercell, where a vacancy is introduced by randomly deleting an Fe atom. One hydrogen atom is randomly placed at the T- or O-sites in the matrix. 94 structures are included in the dataset. The DFT energy and charge density calculations were performed without structural relaxation.

\textit{bcc-Fe--H GB} dataset contains multiple $\Sigma3$$\langle 110 \rangle$ grain boundary (GB) structures in bcc-Fe. The atomic structure was taken from Ref.~\cite{Nakashima2000}. First, the atomic positions and lattice of the GB structures were relaxed. Then, a hydrogen atom was inserted at each Voronoi vertex near the GB without further structural relaxation. A total of 52 configurations were created. Each configuration contains 13--37 atoms.

\item \textit{bcc/fcc/hcp-Fe--C--H interstitial solid solutions} datasets contain 112/49/75 atomic configurations for bcc/fcc/hcp, respectively. The hydrogen and carbon atoms are randomly placed at the T- or O-sites in the matrix. Supercells of $3a_{(\cdot)}\times3a_{(\cdot)}\times3a_{(\cdot)}$ for bcc/fcc and $3a_{\text{hcp}}\times3a_{\text{hcp}}\times3c_{\text{hcp}}$ for hcp. ($a_\text{fcc}=3.66$ \AA, $a_\text{hcp}=2.45$ \AA, and $c_\text{hcp}=3.88$ \AA.)

\item \textit{bcc/fcc/hcp-Fe--C--H substitutional solid solutions} datasets contain 279/45/30 atomic configurations for bcc/fcc/hcp, respectively. Each including one interstitially embedded hydrogen and 1--5 substitutionally embedded carbon atoms. The hydrogen atom is randomly placed at the T- or O-sites in the $3a_{(\cdot)}\times3a_{(\cdot)}\times3a_{(\cdot)}$ for bcc/fcc and $3a^{\prime}_{\text{hcp}}\times3a^{\prime}_{\text{hcp}}\times3c^{\prime}_{\text{hcp}}$ for hcp. ($a_\text{hcp}=2.38$ \AA, and $c^{\prime}_\text{hcp}=3.80$ \AA.)
\end{itemize}

\clearpage

\section{Training curves}
\label{sec:training_curves}

\begin{figure}[hp]
\centering
\includegraphics[width=\textwidth]{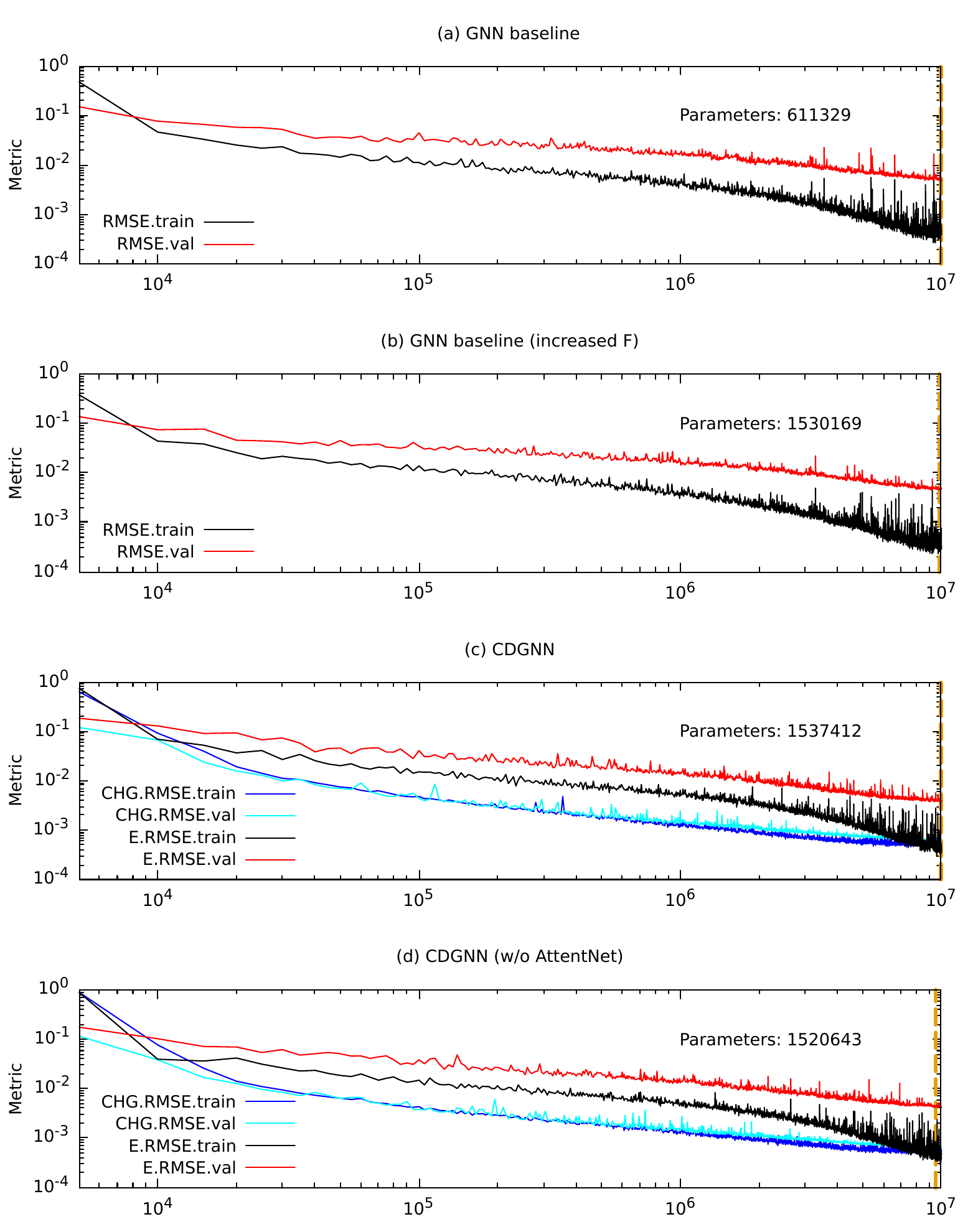}
\caption{Training curves for equivariant GNN and CDGNN models using the \textit{QM9under14atoms} dataset ($M$ = 15862). The yellow dashed lines indicate the best models.
}
\label{fig:learining-curve-total-QM9}
\end{figure}

\begin{figure}[hp]
\centering
\includegraphics[width=\textwidth]{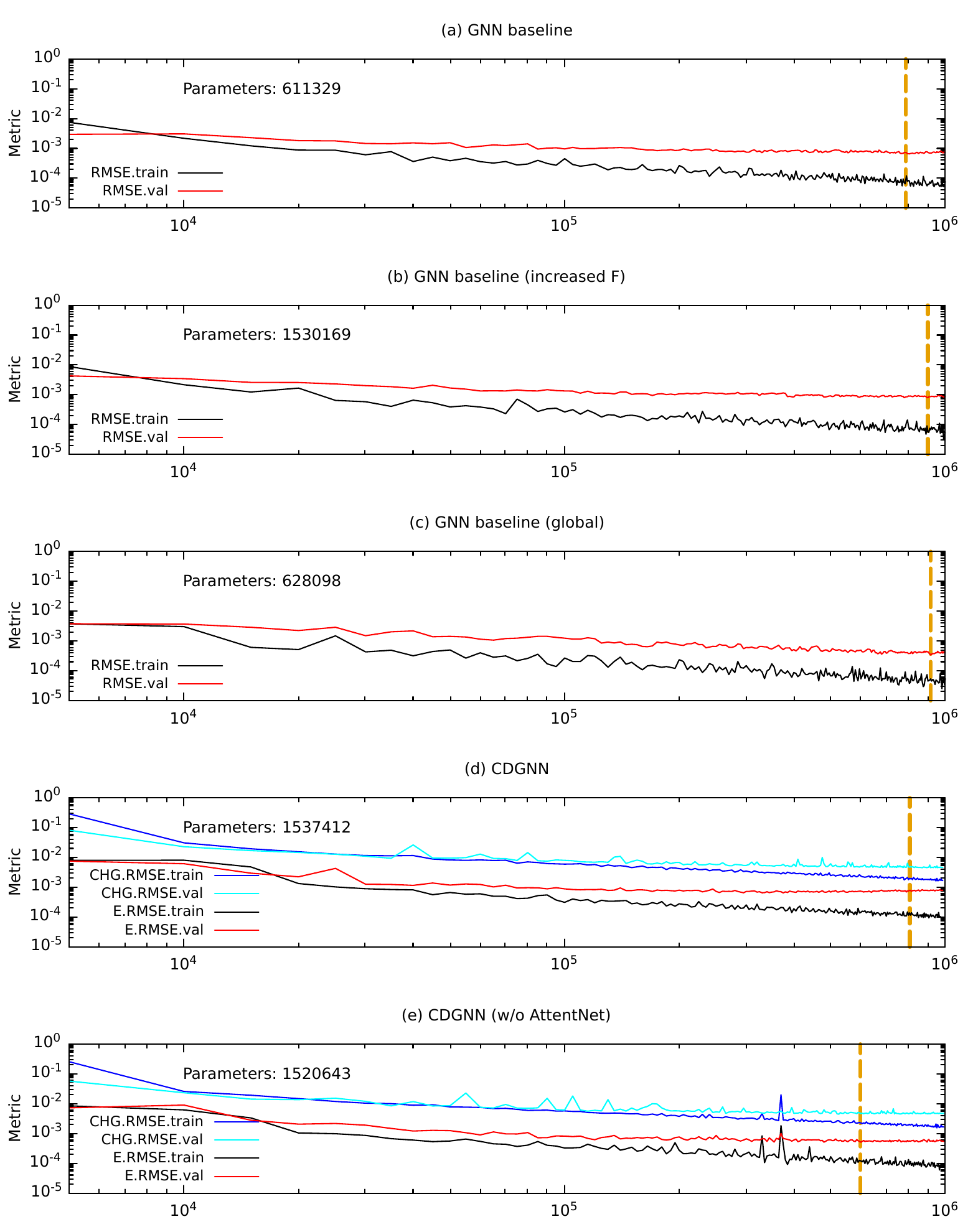}
\caption{Training curves for equivariant GNN and CDGNN models using the \textit{carbide--H} dataset ($M$ = 657). The yellow dashed lines indicate the best models.}
\label{fig:learning_curve_total}
\end{figure}

\clearpage

\section{Evaluation of energy prediction}
\label{sec:energy_prediction}

\begin{figure*}[h]
\centering
\includegraphics[width=\textwidth]{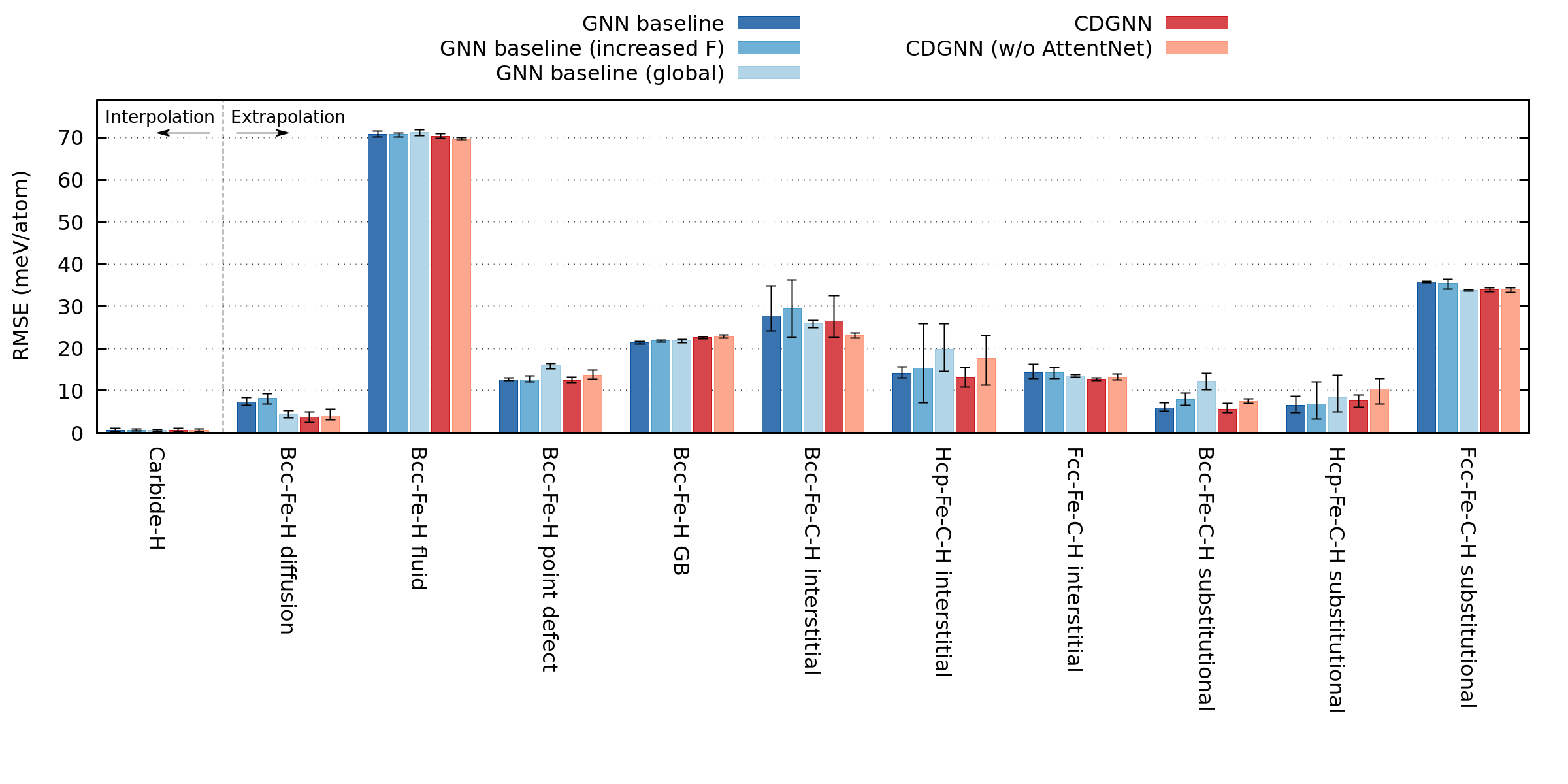}
\caption{Energy RMSE of various models trained on the \textit{carbide--H} dataset for predicting H segregation energy. The \textit{carbide--H} results shown here correspond to the interpolation test split, which was not included in the training process. GNN baseline (global) represents the baseline model with an additional global regressor.}
\label{fig:rmse-Fe-H_glb}
\end{figure*}

\begin{figure}[hp]
\centering
\includegraphics[width=.5\textwidth]{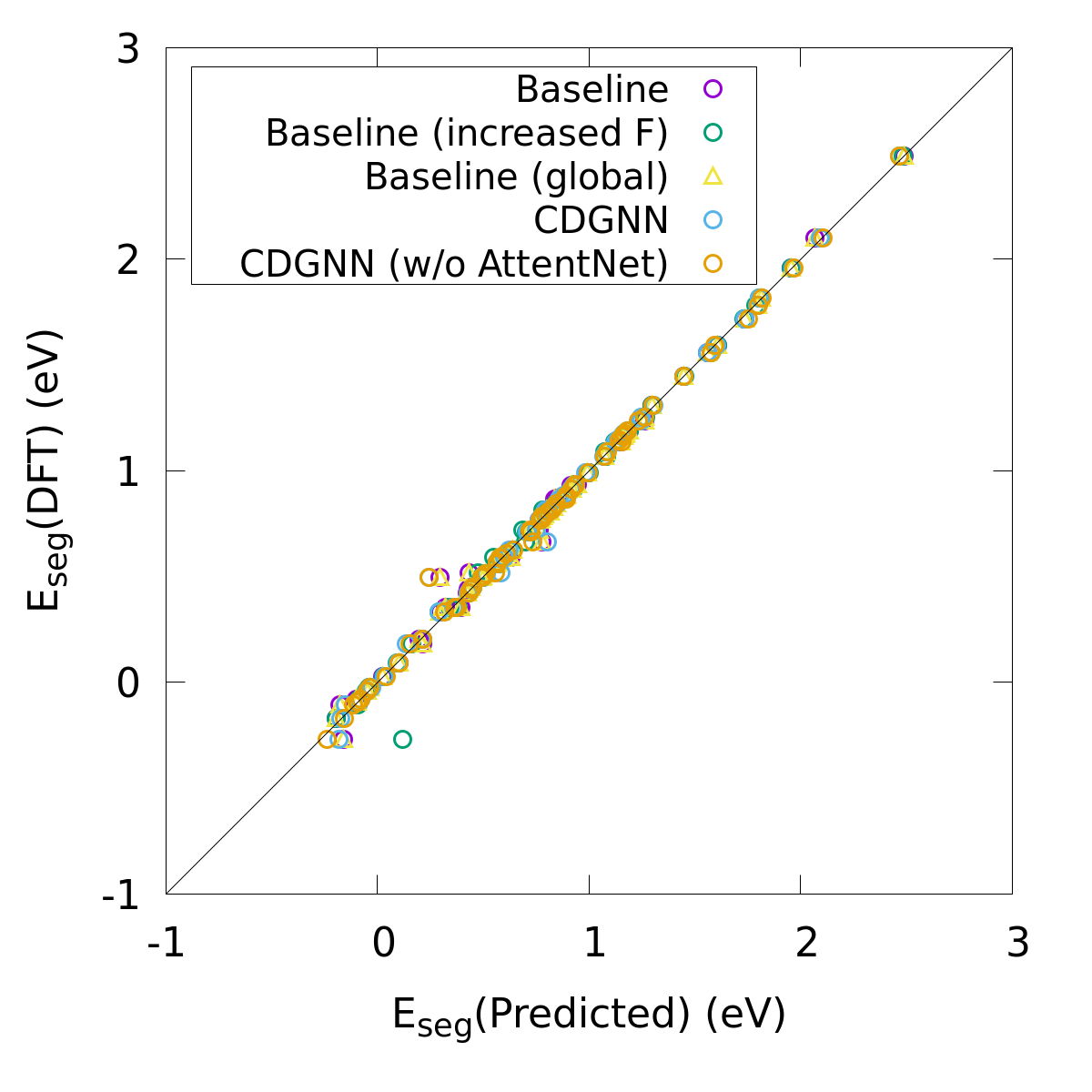}
\caption{Hydrogen segregation energy ($E_\text{seg}$) obtained by DFT calculations and model predictions on the interpolation set \textit{carbide--H}. }
\label{fig:e_inter}
\end{figure}

\begin{figure*}[hp]
\centering
\includegraphics[width=\textwidth]{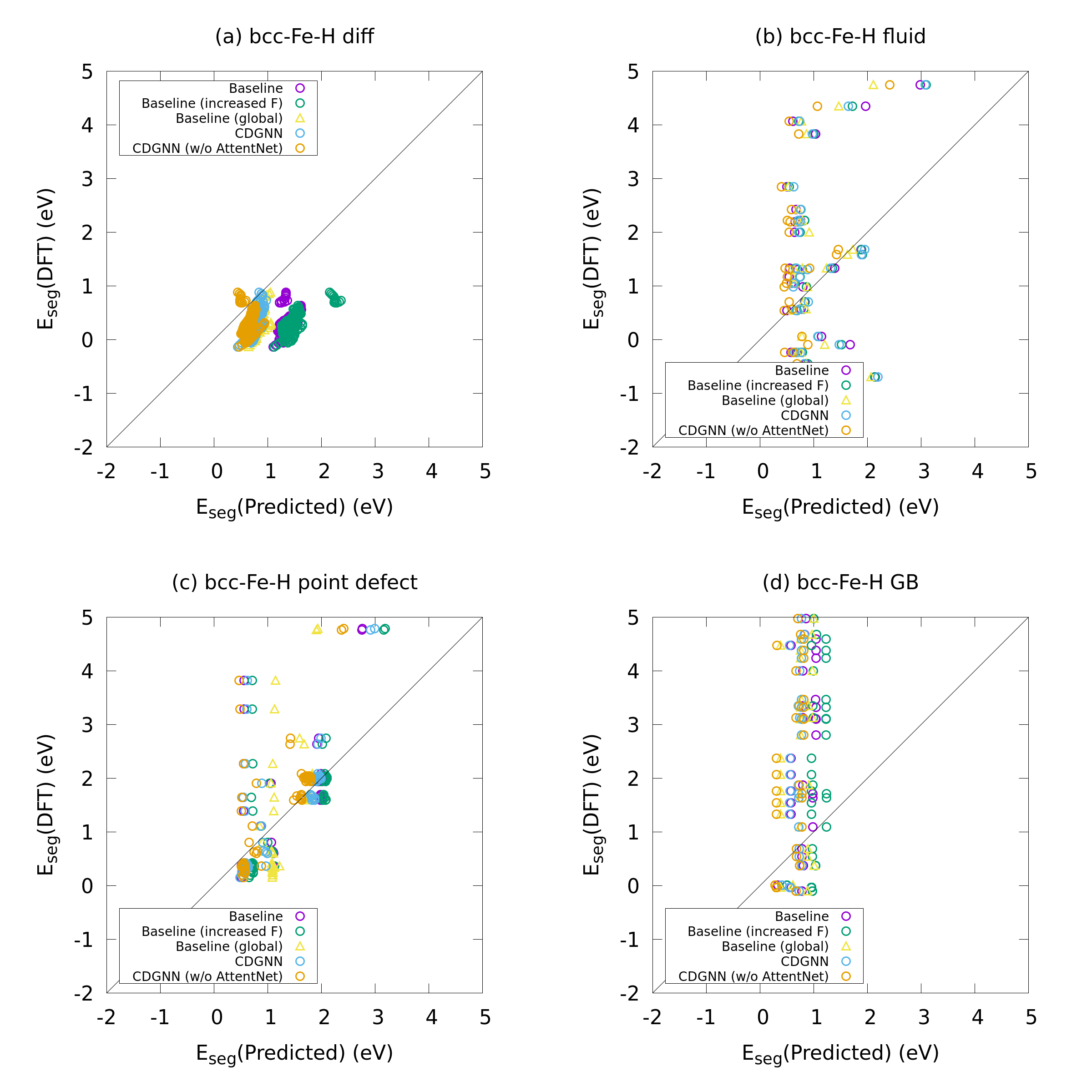}
\caption{Hydrogen segregation energy ($E_\text{seg}$) obtained by DFT calculations and model predictions using extrapolation datasets in the Fe--H systems. }
\label{fig:e_extra_1}
\end{figure*}

\begin{figure*}[hp]
\centering
\includegraphics[width=\textwidth]{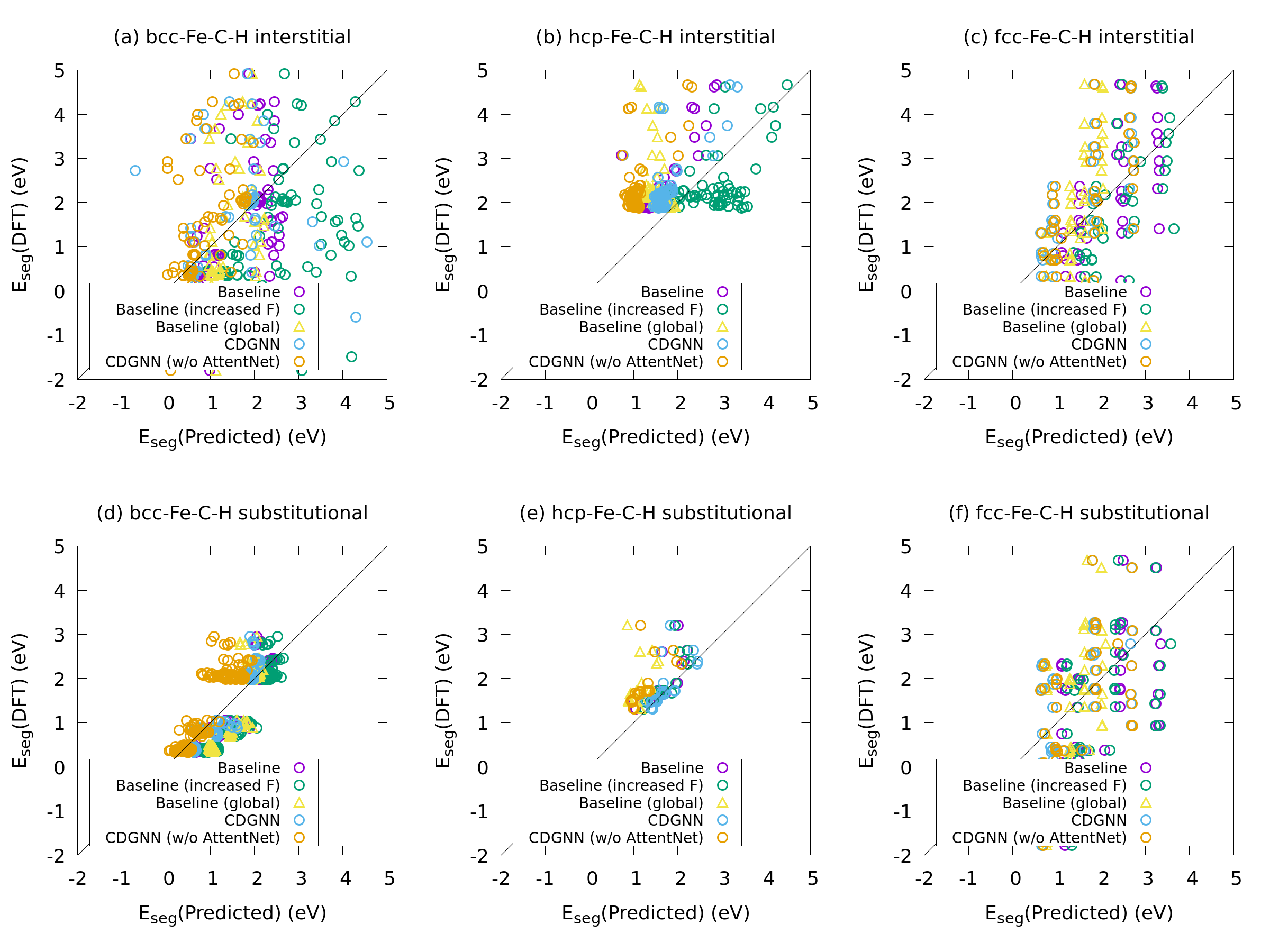}
\caption{Hydrogen segregation energy ($E_\text{seg}$) obtained by DFT calculations and model predictions using additional extrapolation datasets in the Fe--C--H systems. }
\label{fig:e_extra_2}
\end{figure*}


\clearpage

\section{Training and test results of the CDGNN model with a frozen atom branch}
\label{sec:freeze}

\vspace{10 mm}
\begin{figure}[hp]
\centering
\includegraphics[width=\textwidth]{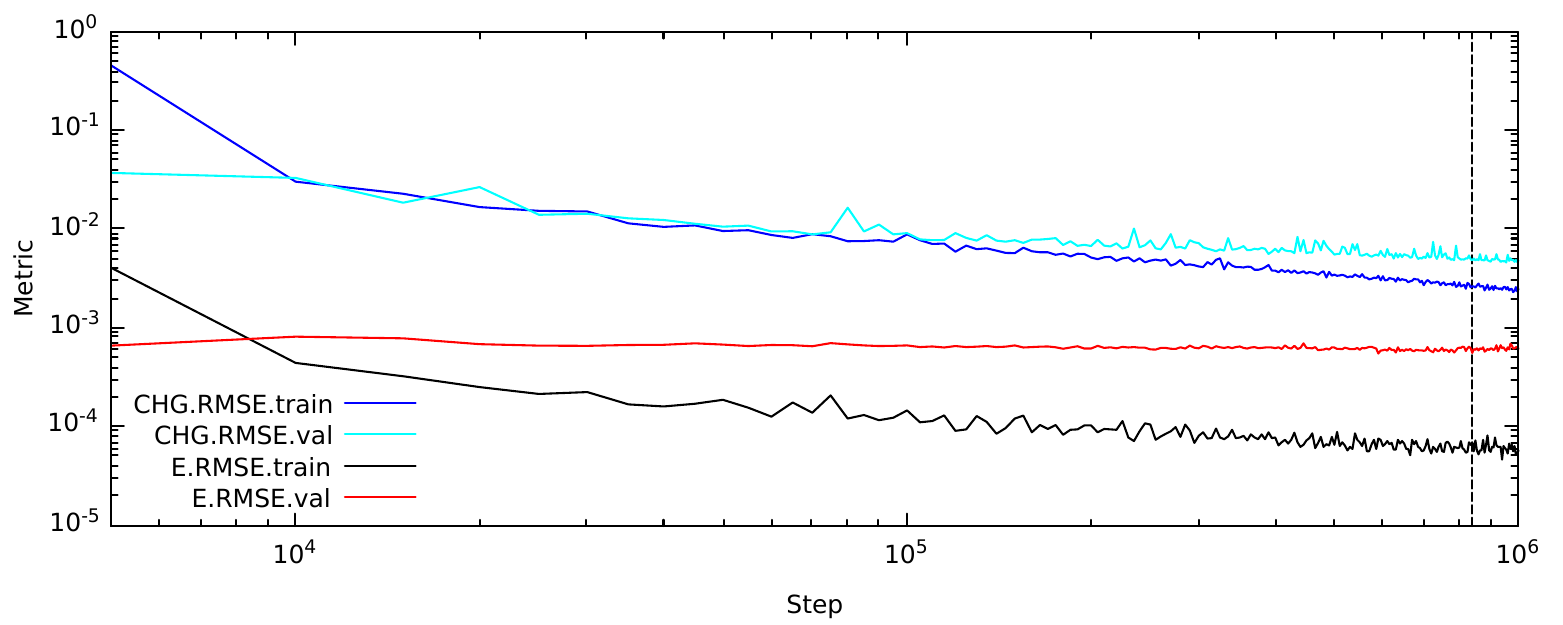}
\caption{Training curves for the CDGNN model with a frozen atom branch using the \textit{carbide--H} dataset. The vertical dashed lines indicate the best models.}
\label{fig:learning_curve_freeze}
\end{figure}

\begin{figure}[hp]
\centering
\includegraphics[width=\textwidth]{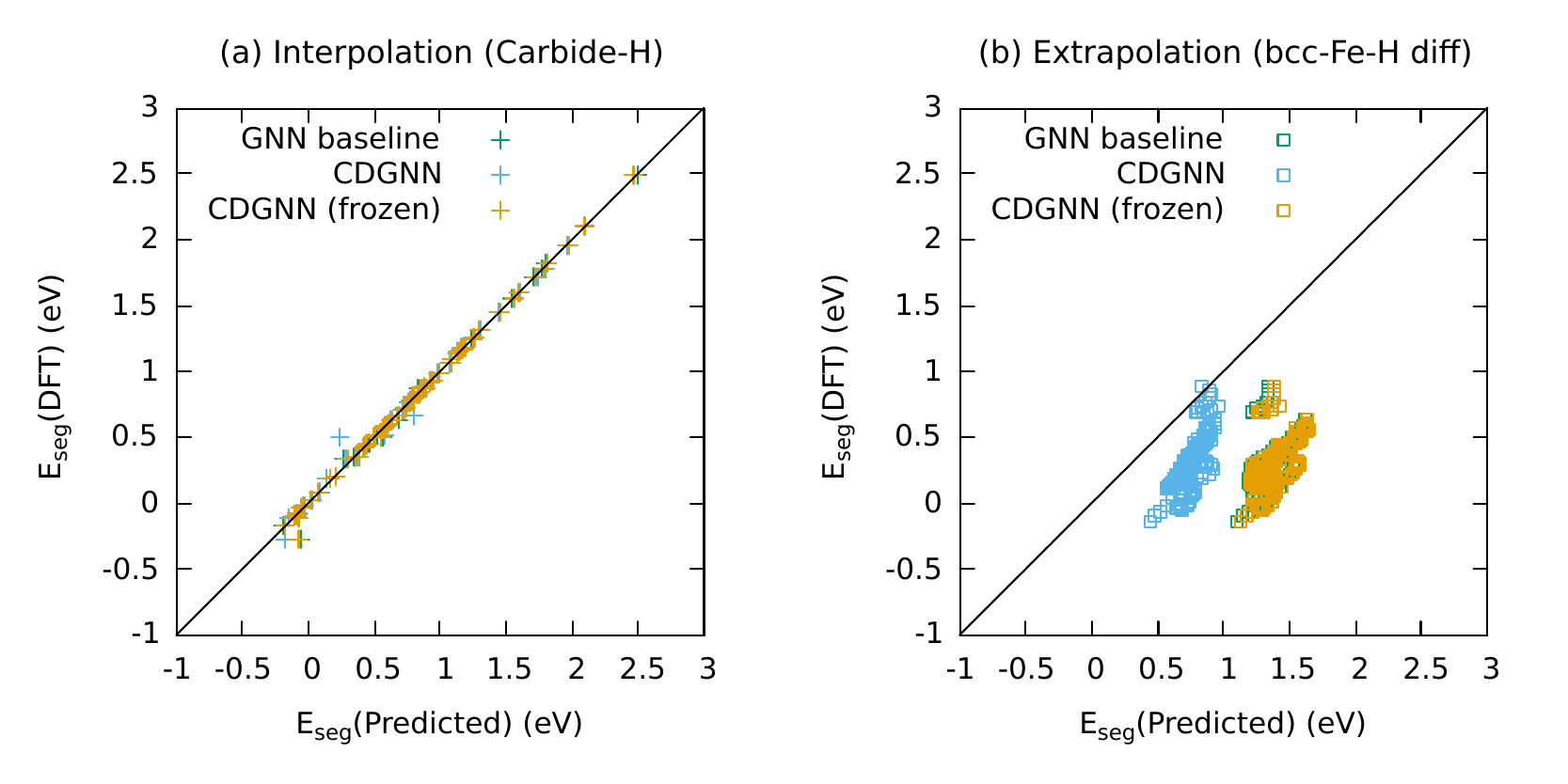}
\caption{Test results of the CDGNN model trained with frozen atom branch on the interpolation and an extrapolation dataset.}
\label{fig:freeze}
\end{figure}

\clearpage

\section{Evaluation of charge density prediction}
\label{sec:deepdft}

\begin{figure}[hp]
\centering
\includegraphics[width=\textwidth]{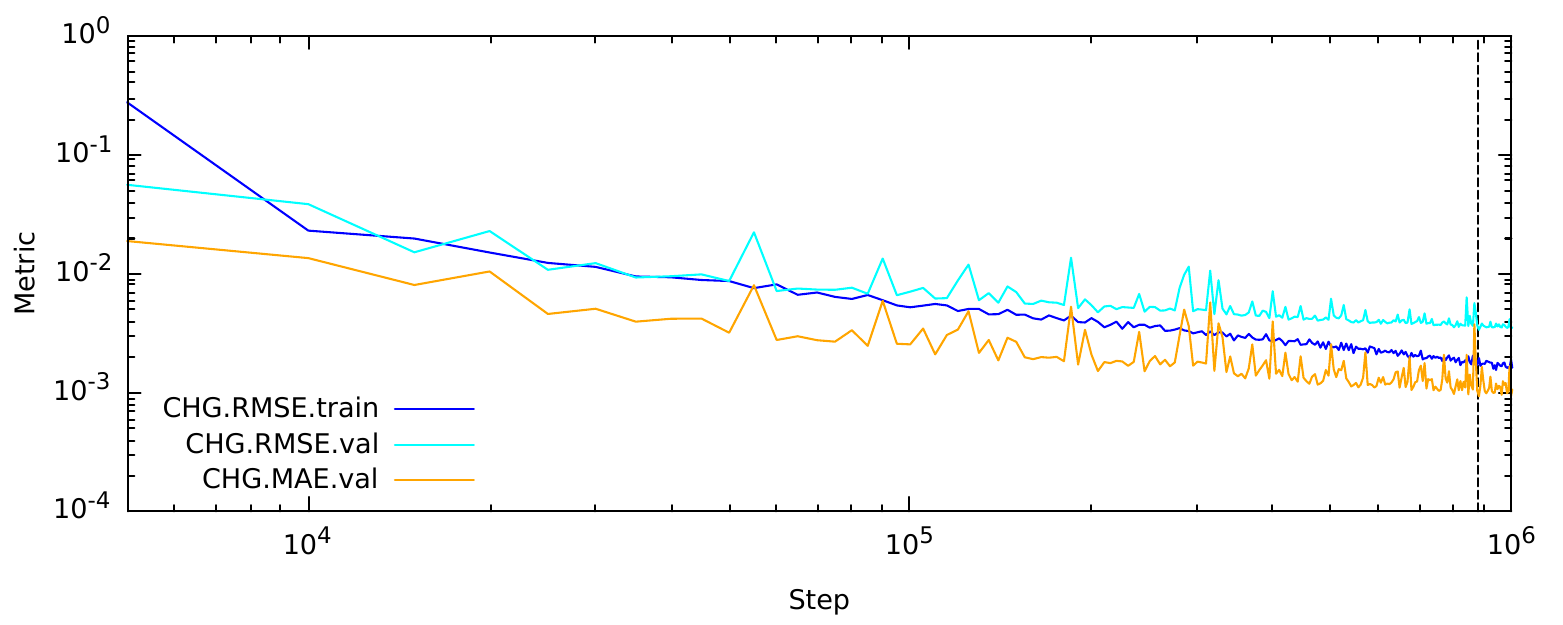}
\caption{Training curves for equivariant DeepDFT model using the \textit{carbide--H} dataset. The vertical dashed lines indicate the best models, which were selected based on $\text{MAE}_\text{density}$ of the validation set. A feature size $F$ = 128, a cutoff radius $r_{c}$ = 6~\AA, and interaction steps $T$ = 3 were used.}
\label{fig:learning_curve_DeepDFT}
\end{figure}

\begin{figure}[hp]
\centering
\includegraphics[width=\textwidth]{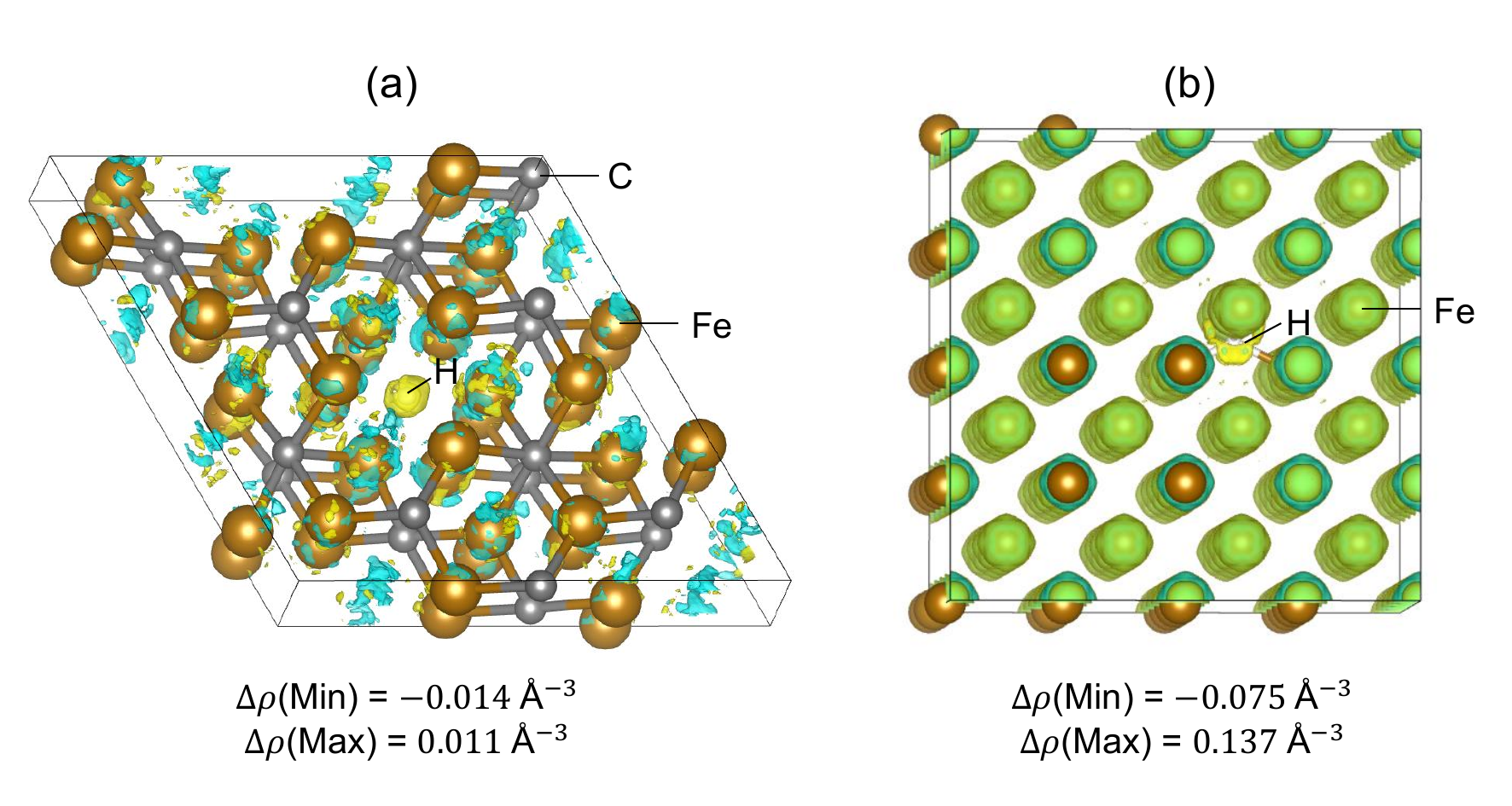}
\caption{Difference of charge densities predicted by CDGNN and equivariant DeepDFT models for (a) H-embedded $\varepsilon$-carbide in the interpolated \textit{carbide--H} set and (b) H-embedded bcc-Fe in the extrapolated \textit{bcc-Fe--H diffusion} set. The charge density difference was calculated as $\Delta\rho = \rho (\text{CDGNN}) - \rho (\text{DeepDFT})$ within the same 3D-grid with grid step of 0.05~\AA. The charge density difference was plotted with isosurface of 0.003~\AA$^{-3}$. The yellow and blue regions represent positive and negative values, respectively.}
\label{fig:chg_diff}
\end{figure}

\begin{figure*}[pt]
\centering
\includegraphics[width=\textwidth]{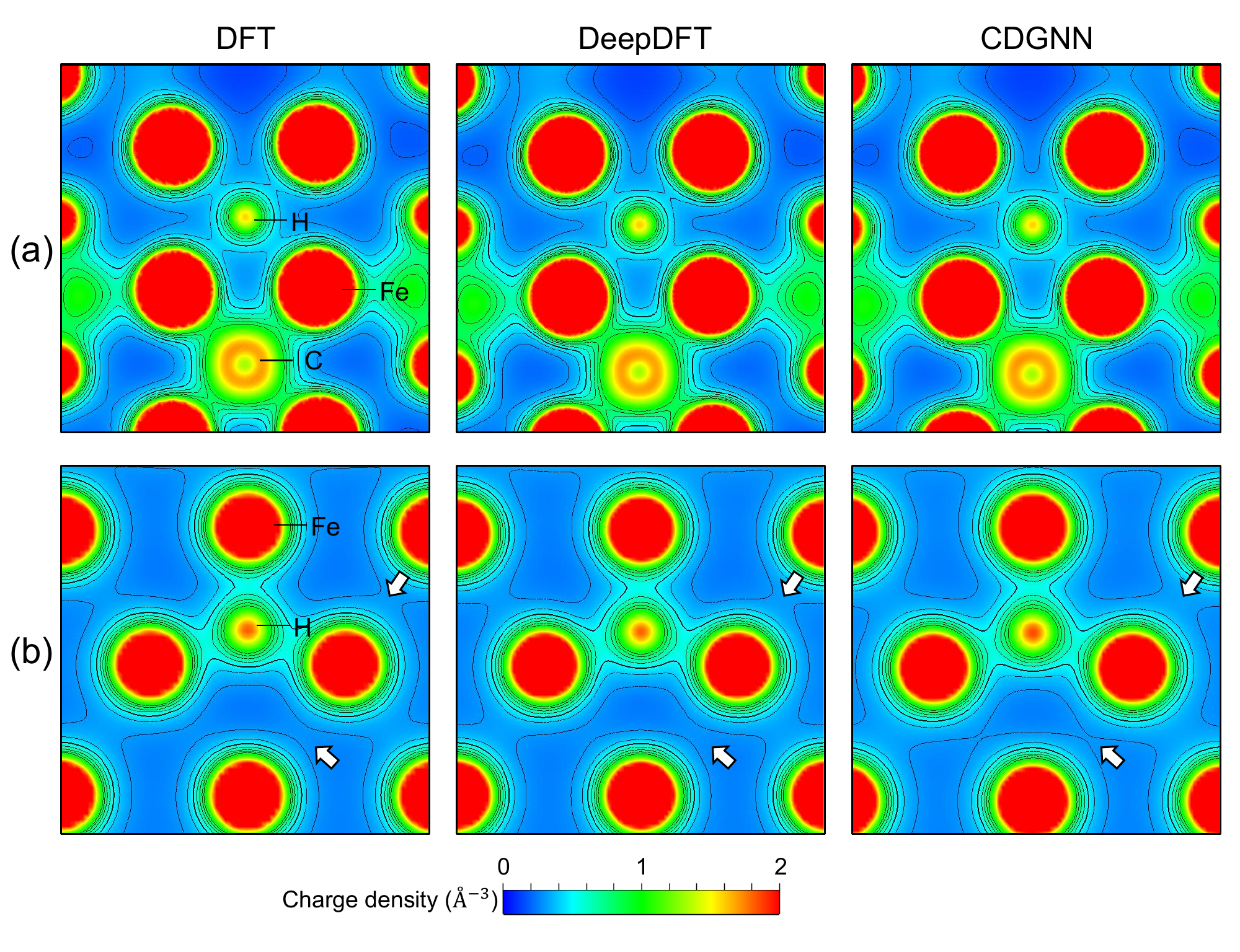}
\caption{2D slices of the charge density map calculated by DFT and predicted by the equivariant DeepDFT and the CDGNN models: (a) $\varepsilon$-carbide with hydrogen-embedded at the octahedral site in interpolation set and (b) bcc-Fe with H-embedded at the tetrahedral site in extrapolation set. The CDGNN prediction aligns well with the DFT and DeepDFT results. For (b) in the extrapolation set, the CDGNN model slightly overestimate the charge density values between the two Fe atoms, as indicated by the arrows.}
\label{fig:chg_2d}
\end{figure*}

\clearpage

\section{Sensitivity to the weights of the loss function}
\label{sec:sensitivity_to_weights}

\begin{figure}[hp]
\centering
\includegraphics[width=\textwidth]{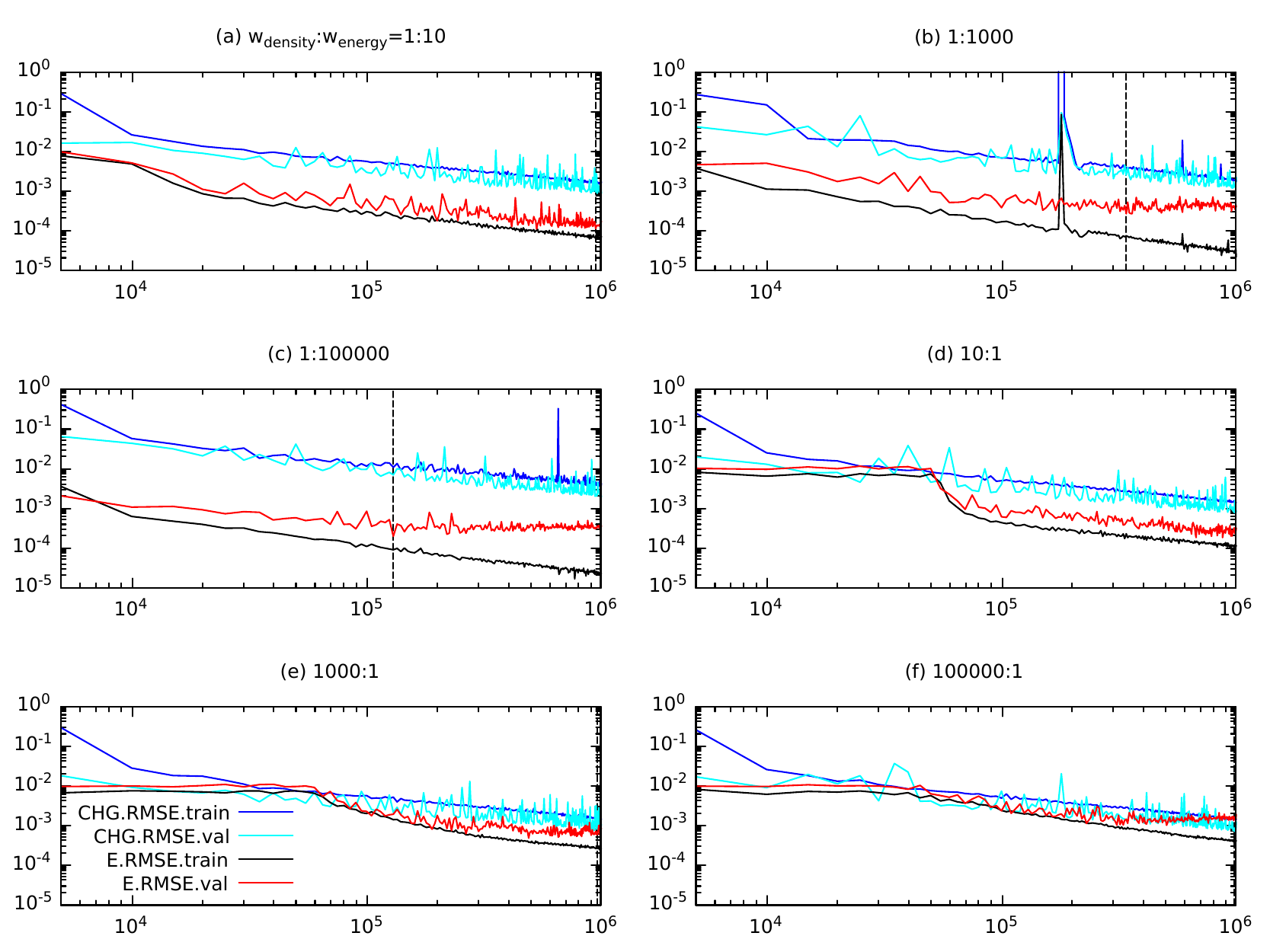}
\caption{Training curves of the CDGNN models using different settings of $w_\text{density} : w_\text{energy}$ of the loss function. The vertical dashed lines indicate the best models.}
\label{fig:training_curve_CDG_TUNE}
\end{figure}

\begin{figure}[hp]
\centering
\includegraphics[width=\textwidth]{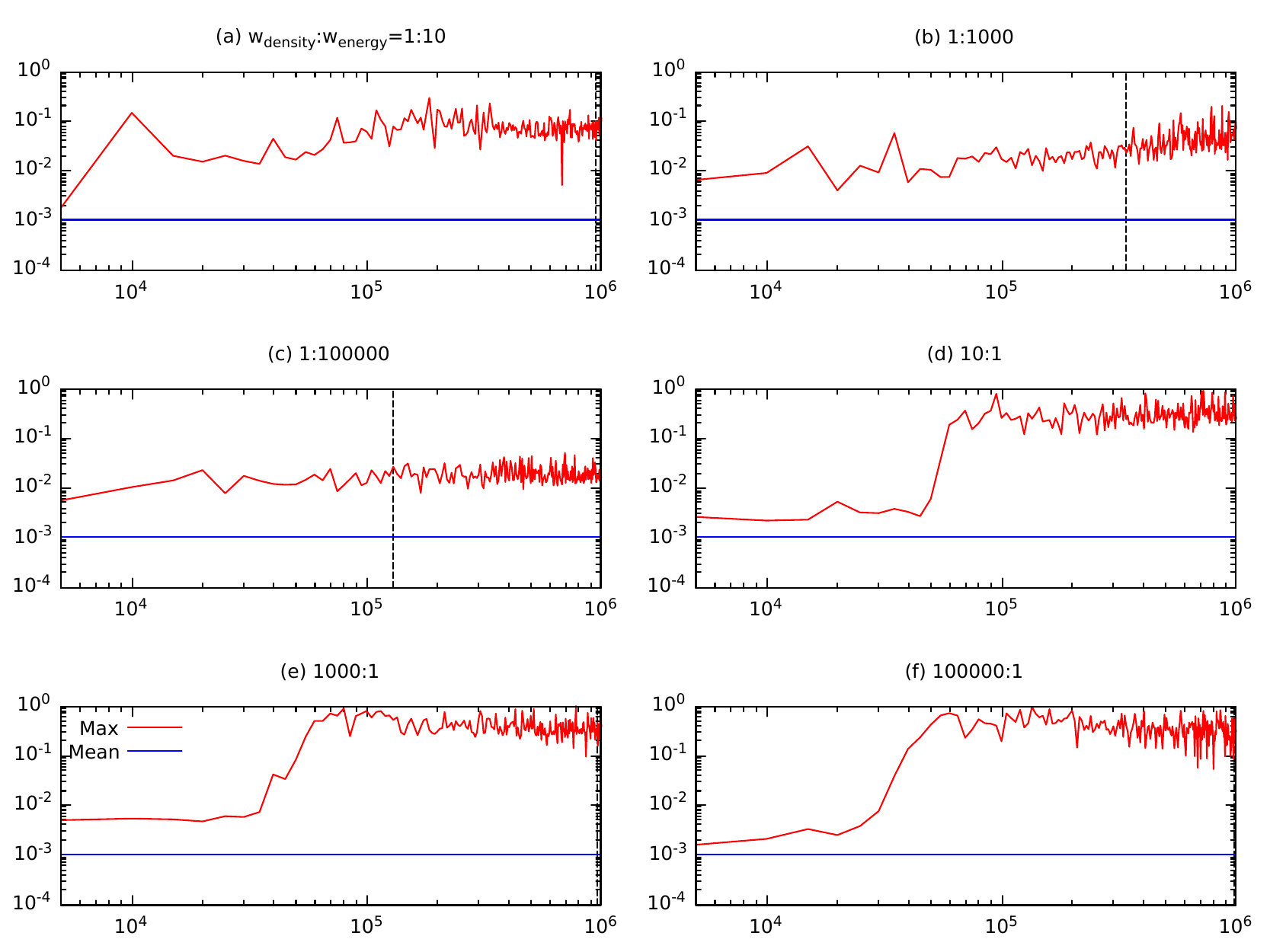}
\caption{Important scores obtained by the attention mechanism during the training of the CDGNN models using different settings of $w_\text{density} : w_\text{energy}$. The vertical dashed lines indicate the best models.}
\label{fig:attent_CDG_TUNE}
\end{figure}

\begin{figure}[hp]
\centering
\includegraphics[width=\textwidth]{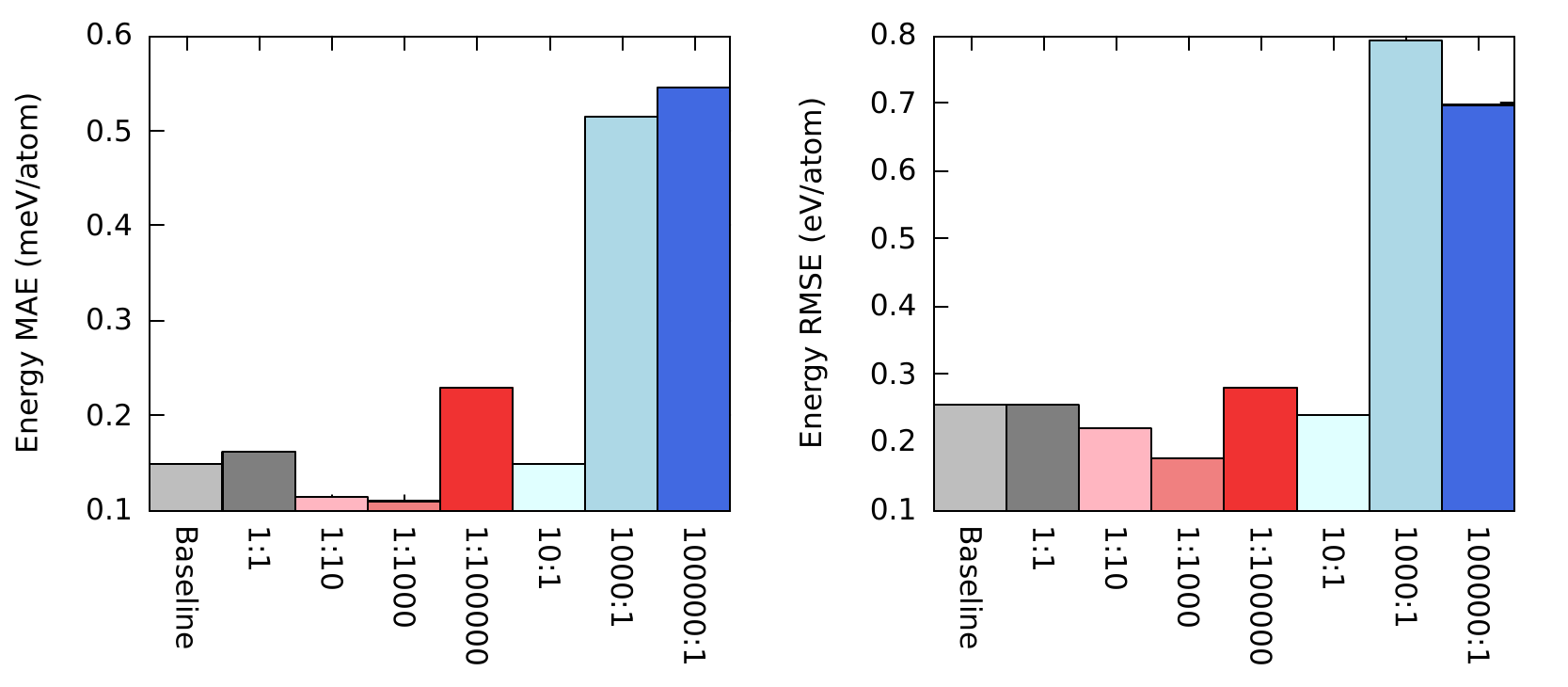}
\caption{Energy MAE and RMSE on the interpolation set, using various CDGNN models trained with different settings of $w_\text{density} : w_\text{energy}$. The baseline is the geometry-only equivariant GNN model.}
\label{fig:inter_CDG_TUNE}
\end{figure}

To estimate the influence of the charge density ($w_\text{density}$) and energy ($w_\text{energy}$) weights of the loss function on training, multiple CDGNN models were trained using different weight settings. A subset of the \textit{carbide--H} dataset containing 356 configurations was used and divided into training, validation, and interpolation test sets at a ratio of $8:1:1$. The training curves are shown in Fig.~\ref{fig:training_curve_CDG_TUNE}. The results of important scores are shown in \ref{fig:attent_CDG_TUNE}. The interpolation test of the energy prediction is shown in \ref{fig:inter_CDG_TUNE}. 

Figure~\ref{fig:training_curve_CDG_TUNE} shows that increasing the weight of one parameter reduces its training error. Conversely, the training error of the other parameter increases.  When $w_\text{density}$ and $w_\text{energy}$ differ by a smaller amount (e.g., a ratio of $1:10$), the charge density branch has a greater loss. This suggests that the charge density part is more difficult to learn. Additionally, increasing the share of $w_\text{energy}$ does not result in a significant reduction in the energy loss.
Results in Fig.~\ref{fig:attent_CDG_TUNE} show that increasing the share of $w_\text{density}$ may results in the enhancement of the attention scoring mechanism, as the important score reaches its maximum value in fewer training steps.
Using a high share of $w_\text{energy}$ may weaken this function.
As shown in Fig.~\ref{fig:inter_CDG_TUNE}, the energy prediction accuracy on the interpolation set indicates that the CDGNN model trained using $w_\text{density} : w_\text{energy}$ ratios of $1:10$, $1:1000$, or $10:1$ achieves higher energy prediction accuracy on the interpolation set than the baseline GNN model.  Notably, the accuracy of the energy prediction decreases significantly when the charge density is given a very high weight, such as $1000:1$ or $100000:1$.

\vspace{10 mm}

\end{document}